\documentclass[a4paper,twocolumn,10pt]{article}

\usepackage{graphicx}
\usepackage{epsfig}
\usepackage{balance}
\usepackage{subfigure}
\usepackage{times}

\newcommand{\newspacing}{\baselineskip=1.0\normalbaselineskip}

\begin{document}
%
%

\title{AIS for Misbehavior Detection in Wireless Sensor Networks: Performance and Design Principles}

\author{Martin Drozda~~~~~~~Sven Schaust~~~~~~~Helena Szczerbicka\\Leibniz University of Hannover, Department of
  Computer Science\\
 FG Simulation und
Modellierung, Welfengarten 1, 30167 Hannover, Germany\\Email: \{drozda,svs,hsz\}@sim.uni-hannover.de}

\date{}


\maketitle


\newspacing
\begin{abstract} 
A sensor network is a collection of wireless devices that are able to monitor physical or environmental conditions. These devices (nodes) are expected to operate autonomously, be battery powered and have very limited computational capabilities. This makes the task of protecting a sensor network against misbehavior or possible malfunction a challenging problem. In this document we discuss performance of Artificial immune systems (AIS) when used as the mechanism for detecting misbehavior. 

We show that (i) mechanism of the AIS have to be carefully applied in order to avoid security weaknesses, (ii) the choice of genes and their interaction have a profound influence on the performance of the AIS, (iii) randomly created detectors do not comply with limitations imposed by communications protocols and (iv) the data traffic pattern seems not to impact significantly the overall performance. 

We identified a specific MAC layer based gene that showed to be especially useful for detection; genes measure a network's  performance from a node's viewpoint. Furthermore, we identified an interesting complementarity property of genes; this property exploits the local nature of sensor networks and moves the burden of excessive communication from normally behaving nodes to misbehaving nodes.
These results have a direct impact on the design of AIS for sensor networks and on engineering of sensor networks. 
\end{abstract}

\section{Introduction and Motivation}

Sensor networks~\cite{karl2005paa} can be described as a collection of wireless devices with limited computational abilities which are, due to their ad-hoc communication manner, vulnerable to misbehavior and malfunction. It is therefore necessary to support them with a simple, {\em computationally friendly} protection system. 

Due to the limitations of sensor networks, there has been an on-going interest in providing them with a protection solution that would fulfill several basic criteria. The first criterion is the ability of self-learning and self-tuning. Because maintenance of ad hoc networks by a human operator is expected to be sporadic, they have to have a built-in {\em autonomous} mechanism for identifying user behavior that could be potentially damaging to them. This learning mechanism should itself minimize the need for a human intervention, therefore it should be self-tuning to the maximum extent. It must also be computationally conservative and meet the usual condition of high detection rate. The second criterion is the ability to undertake an action against one or several misbehaving users. This should be understood in a wider context of co-operating wireless devices acting in collusion in order to suppress or minimize the adverse impact of such misbehavior. Such a co-operation should have a low message complexity because both the bandwidth and the battery life are of scarce nature. The third and last criterion requires that the protection system does not itself introduce new weaknesses to the systems that it should protect.

An emerging solution that could facilitate implementation of the above criteria are Artificial immune systems (AIS). AIS are based on principles adapted from the Human immune system (HIS)~\cite{janewayjr2001isw,banchereau2000idc,hofmeyr1999ida}; the basic ability of HIS is an efficient detection of potentially harmful foreign agents (viruses, bacteria, etc.). The goal of AIS, in our setting, is the identification of nodes with behavior that could possibly negatively impact the stated mission of the sensor network. 

One of the key design challenges of AIS is to define a suitable set of efficient genes. Genes form a basis for deciding whether a node misbehaves. They can be characterized as measures that describe a network's performance from a node's viewpoint.  Given their purpose, they must be easy to compute and robust against deception.

Misbehavior in wireless sensor networks can take upon different forms: packet dropping, modification of data structures important for routing, modification of packets, skewing of the network's topology or creating ficticious nodes (see~\cite{drozda2006ais} for a more complete list). The reason for sensors (possibly fully controlled by an attacker) to execute any form of misbehavior can range from the desire to save battery power to making a given wireless sensor network non-functional. Malfunction can also be considered a type of unwanted behavior. 


\section{Artificial Immune Systems}

\subsection{Background}

The Human immune system is a rather complex mechanism able to protect humans against an amazing set of extraneous attacks. This system is remarkably efficient, most of the time, in discriminating between {\em self} and {\em non-self} antigens.\footnote{Self and non-self in short.} A non-self antigen is anything that can initiate an immune response; examples are a virus, bacteria, or splinter. The opposite to non-self antigens are self antigens; self antigens are human organism's own cells.

\begin{figure}[!!!ht]
\begin{center}
  {
    \epsfig{file=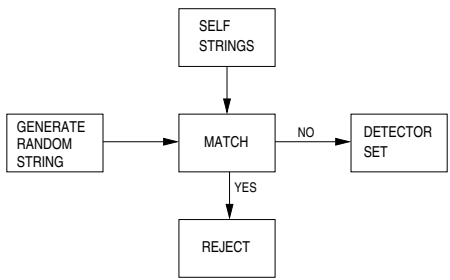, width=.75\linewidth}
  }
\caption{T-cell (detector) generation by random-generate-and-test process. A (bit) string representation is assumed.
}
\label{fig:det_gen}
\end{center}
\end{figure}

\begin{figure}[!!!ht]
\begin{center}
  {
    \epsfig{file=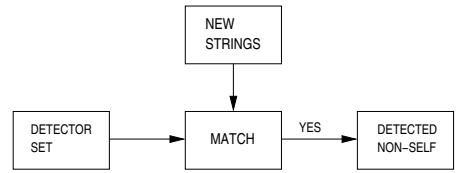, width=.75\linewidth}
  }
\caption{Recognizing non-self is done by matching T-cells (detectors) with suspected non-self antigens (new strings).}
\label{fig:non_self_det}
\end{center}
\end{figure}

\subsection{Learning}

The process of T-cells maturation in thymus is used as an inspiration for learning in AIS. 
The maturation of T-cells (detectors) in thymus is a result of a pseudo-random process. After a T-cell is created (see Fig.~\ref{fig:det_gen}), it undergoes a censoring process called {\em negative selection}. During negative selection T-cells that bind self are destroyed. Remaining T-cells are introduced into the body. The recognition of non-self is then done by simply comparing T-cells that survived negative selection with a suspected non-self. This process is depicted in Fig.~\ref{fig:non_self_det}. It is possible that the self set is incomplete, while a T-cell matures (tolerization period) in the thymus. This could lead to producing T-cells that should have been removed from the thymus and can cause an autoimmune reaction, i.e. it leads to {\em false positives}. 

A deficiency of the negative selection process is that alone it is not sufficient for assessing the damage that a non-self antigen could cause. For example, many bacteria that enter our body are not harmful, therefore an immune reaction is not necessary. T-cells, actors of the adaptive immune system, require co-stimulation from the innate immune system in order to start acting. The innate immune system is able to recognize the presence of harmful non-self antigens and tissue damage, and signal this to certain actors of the adaptive immune system.

The random-generate-and-test approach for producing T-cells (detectors) described above is analyzed in~\cite{dhaeseleer1996iac}. In general, the number of candidate detectors to the self set size needs to be exponential (if a matching rule with fixed matching probability is used). Another problem is a consistent underfitting of the non-self set; there exist ``holes'' in the non-self set that are undetectable. In theory, for some matching rules, the number of holes can be very unfavorable~\cite{stibor2005nsa}. In practical terms, the effect of holes depends on the characteristics of the non-self set, representation and matching rule~\cite{gonzalez2003ebm}. The advantage of this algorithm is its {\em simplicity} and good experimental results in cases when the number of detectors to be produced is fixed and small~\cite{sarafijanovic2004ais}. 
A review of other approaches to detector computation can be found in~\cite{aickelin4isa}.

\section{Sensor Networks}

A sensor network can be defined in graph theoretic
framework as follows: a sensor network is a net $N = (n(t), e(t))$
where $n(t), e(t)$ are the set of nodes and edges at time $t$,
respectively. Nodes correspond to sensors 
 that wish to communicate with each other. An edge between two nodes
$A$ and $B$ is said to exist when $A$ is within the radio transmission
range of $B$ and vice versa. The imposed symmetry of edges is a
usual assumption of many mainstream protocols. The change in the
cardinality of sets $n(t), e(t)$ can be caused by switching on/off one of the sensors, failure, malfunction, removal, signal propagation,
link reliability and other factors. 

Data exchange in a point-to-point (uni-cast) scenario usually proceeds
as follows: a user initiated data exchange leads to a route query at
the network layer of the OSI stack. A routing protocol at that layer
attempts to find a route to the data exchange destination. This
request may result in a path of non-unit length. This means that a data
packet in order to reach the destination has to rely on successive
forwarding by intermediate nodes on the path. An example of an on-demand routing protocol often used in sensor networks is DSR~\cite{johnson1996dsr}. Route search in this protocol is started only when a route to a destination is needed. This is done by flooding the network with RREQ\footnote{RREQ = Route Request, RREP = Route Reply, RERR = Route Error.} control packets. The destination node or an intermediate node that knows a route to the destination will reply with a RREP control packet. This RREP follows the route back to the source node and updates routing tables at each node that it traverses. A RERR packet is sent to the connection originator when a node finds out that the next node on the forwarding path is not replaying.

At the MAC layer of the OSI protocol stack, the medium reservation is often contention based. In order to transmit a data packet, the IEEE 802.11 MAC protocol uses carrier sensing with an RTS-CTS-DATA-ACK handshake.\footnote{RTS = Ready to send, CTS = Clear to send, ACK = Acknowledgment.} Should the medium not be available or the handshake fails, an exponential back-off algorithm is used. This is combined with a mechanism that makes it easier for neighboring nodes to estimate transmission durations. This is done by exchange of duration values and their subsequent storing in a data structure known as Network allocation vector (NAV). With the goal to save battery power, researchers suggested, a sleep-wake-up schedule for nodes would be appropriate. This means that nodes do not listen continuously to the medium, but switch themselves off and wake up again after a predetermined period of time. Such a sleep and wake-up schedule is similarly to duration values exchanged among nodes. An example of a MAC protocol, designed specifically for sensor networks, that uses such a schedule is the S-MAC~\cite{ye2004mac}. A sleep and wake-up schedule can severely limit operation of a node in {\em promiscuous mode}. In promiscuous mode, a node listens to the on-going traffic in the neighborhood and collects information from the overheard packets. This technique is used e.g. in DSR for improved propagation of routing information.

Movement of nodes can be modeled by means of a mobility model. A well-known mobility model is the {\em Random waypoint model}~\cite{johnson1996dsr}. 
In this model, nodes move from the current position to a new randomly
generated position at a predetermined speed. After reaching the new
destination a new random position is computed. Nodes pause at the current position for a time period $t$ before moving to the new random position. 

For more information on sensor networks, we refer the reader to~\cite{karl2005paa}.

\section{Summary of Results}

Motivated by the positive results reported in~\cite{hofmeyr1999ida,sarafijanovic2004ais} we have undertaken a detailed performance study of AIS with focus on sensor networks.
The general conclusions that can be drawn from the study presented in this document are:

\smallskip

1.~Given the ranges of input parameters that we used and considering the computational capabilities of current sensor devices, we conclude that AIS based misbehavior detection offers a decent detection rate.

2.~One of the main challenges in designing well performing AIS for sensor networks is the set of ``genes''. This is similar to observations made in~\cite{leboudec2004ais}.

3.~Our results suggest that to increase the detection performance, an AIS should benefit from information available at all layers of the OSI protocol stack; this includes also detection performance with regards to a simplistic flavor of misbehavior such as packet dropping. This supports ideas shortly discussed in~\cite{zhang2003idt} where the authors suggest that information available at the application layer deserves more attention.

4.~
We observed that somewhat surprisingly a gene based purely on the MAC layer significantly contributed to the overall detection performance. This gene poses less limitations when a MAC protocol with a sleep-wake-up schedule such as the S-MAC~\cite{ye2004mac} is used.

5.~It is desirable to use genes that are {\em ``complementary''} with respect to each other. We demonstrated that two genes, one that measures correct forwarding of data packets, and the other one that indirectly measures the medium contention, have exactly this property.  

6.~We only used a single instance of learning and detection mechanism per node. This is different from approach used in~\cite{hofmeyr1999ida,sarafijanovic2004ais}, where one instance was used for each of $m$ possible neighbors. 
Our performance results 
show that the approach in~\cite{hofmeyr1999ida,sarafijanovic2004ais} may not be feasible for sensor networks. It may allow for an easy Sybil attack and, in general, $m = n-1$ instances might be necessary, where $n$ is the total number of sensors in the network. Instead, we suggest that flagging a node as misbehaving should, if possible, be based on detection at several nodes.

7.~Only less than 5\% detectors were used in detecting misbehavior. This suggests that many of the detectors do not comply with constraints imposed by the communications protocols; this is an important fact when designing AIS for sensor networks because the memory capacity at sensors is expected to be very limited.

8.~The data traffic properties seem not to impact the performance. This is demonstrated by similar detection performance, when data traffic is modeled as constant bit rate and Poisson distributed data packet stream, respectively. 

9.~We were unable to distinguish between nodes that misbehave (e.g. deliberately drop data packets) and nodes with a behavior resembling a misbehavior (e.g. drop data packets due to medium contention). This motivates the use of danger signals as described in~\cite{aickelin2003dtl,greensmith2005idc}. The approach applied in~\cite{sarafijanovic2004ais} does, however, not completely fit sensor networks since these might implement only a simplified version of the transport layer.

\section{AIS for Sensor Networks: Design Principles}

In our approach, each node produces and maintains its own set of detectors. This means that we applied a direct one-to-one mapping between a human body with a thymus and a node. We represent self, non-self and detector strings as bit-strings. The matching rule employed is the {\em r-contiguous bits matching rule}. Two bit-strings of equal length match under the r-contiguous matching rule if there exists a substring of length $r$ at position $p$ in each of them and these substrings are identical.  
Detectors are produced by the process shown in Fig.~\ref{fig:det_gen}, i.e. by means of negative selection when detectors are created randomly and tested against a set of self strings. 

Each antigen consists of several genes. {\em Genes} are performance measures that a node can acquire locally without the help from another node. In practical terms this means that an antigen consists of $x$ genes; each of them encodes a performance measure, averaged in our case over a time window. An antigen is then created by concatenating the $x$ genes. 

When choosing the correct genes, the choice is limited due to the simplified OSI protocol stack of sensors. For example, Mica2 sensors~\cite{xbow} 
 using the TinyOS operating system do not guarantee any end-to-end connection reliability (transport layer), leaving only data traffic at the lower layers for consideration. 

Let us assume that the routing protocol finds for a connection the path ${s_s, s_1,...,s_i, s_{i+1},s_{i+2},..., s_d}$ from the source node $s_s$ to the destination node $s_d$, where $s_s \neq s_d$ and $s_{i+1} \neq s_d$. 
We have used the following {\em genes} to capture certain aspects of MAC and routing layer traffic information (we averaged over a time period (window size) of 500 seconds):

\smallskip
\begin{enumerate}
\item[] {\bf MAC Layer:}

\item[\#1] Ratio of complete MAC layer handshakes between nodes $s_i$ and $s_{i+1}$ and RTS packets sent by $s_i$ to $s_{i+1}$. If there is no traffic between two nodes this ratio is set to $\infty$ (a large number). This ratio is averaged over a time period. A complete handshake is defined as a completed sequence of RTS, CTS, DATA, ACK packets between  $s_i$ and $s_{i+1}$.

\item[\#2] Ratio of data packets sent from $s_i$ to $s_{i+1}$ and then subsequently forwarded by $s_{i+1}$ to $s_{i+2}$. If there is no traffic between two nodes this ratio is set to $\infty$ (a large number). This ratio is computed by $s_i$ in promiscuous mode and, as in the previous case, averaged over a time period. This gene was adapted from the watchdog idea in~\cite{marti2000mrm}.

\item[\#3] Time delay that a data packet spends at $s_{i+1}$ before being forwarded to $s_{i+2}$. The time delay is observed by $s_i$ in promiscuous mode. If there is no traffic between two nodes the time delay is set to zero. This measure is averaged over a time period. This gene is a quantitative extension of the previous gene.

\smallskip
\item[] {\bf Routing Layer:}

\item[\#4] The same ratio as in \#2 but computed separately for RERR routing packets.

\item[\#5] The same delay as in \#3 but computed separately for RERR routing packets. 

\end{enumerate}
\smallskip

The Gene \#1 can be characterized as MAC layer quality oriented  -- {\em it indirectly measures the medium contention level}.
The remaining genes are watchdog oriented. This means that they more strictly fit a certain kind of misbehavior. The Gene \#2 can help detect whether packets get correctly forwarded;  the Gene \#3 can help detect whether forwarding of packets does not get intentionally delayed. 
As we will show later, in the particular type of misbehavior (packet dropping) that we applied, the first two genes come out as ``the strongest''. The disadvantage of the watchdog based genes is that due to limited battery power, nodes could operate using a sleep-wake-up schedule similar to the one used in the S-MAC. This would mean that the node $s_i$ has to stay awake until the node  $s_{i+1}$ (monitored node) correctly transmits to $s_{i+2}$. The consequence would be a longer wake-up time and possible restrictions in publishing sleep-wake-up schedules.  

In~\cite{leboudec2004ais} the authors applied a different a set of genes, based only on the DSR routing protocol. The observed set of events was the following: A = RREQ sent, B = RREP sent, C = RERR sent, D = DATA sent and IP source address is not of the monitored (neighboring) node, E = RREQ received, F = RREP received, G = RERR received, H = DATA received and the IP destination address is not of the monitored node. The events D and H take into consideration that the source and destination nodes of a connection might appear as misbehaving as they seem to ``deliberately'' create and delete data packets. Then the set of their four genes is as follows:

\begin{enumerate}
\item[\#1] Number of E over a time period.
\item[\#2] Number of (E*(A or B)) over a time period.
\item[\#3] Number of H over a time period.
\item[\#4] Number of (H*D) over a time period.
\end{enumerate}

The time period (window size) in their case was 10s; * is the Kleene star operator (zero or more occurrences of any event(s) are possible). Similar to our watchdog genes, these genes impose additional requirements on MAC protocols such as the S-MAC. Their dependence on the operation in promiscuous mode is, however, more pronounced as a node has to continuously observe packet events at all monitored  nodes.

The research in the area of what and to what extent can be or should be locally measured at a node, is independent of the learning mechanism used (negative selection in both cases). Performance of an AIS can partly depend on the ordering and the number of used genes. Since longer antigens (consisting of more genes) indirectly imply more candidate detectors, the number of genes should be carefully considered. Given $x$ genes, it is possible to order them in $x!$ different ways. In our experience, the rules for ordering genes and the number of genes can be summed up as follows:

\smallskip
1)~Keep the number of genes small. In our experiments, we show that with respect to the learning mechanism used and the expected deployment (sensor networks), 2-3 genes are enough for detecting a basic type of misbehavior.

\smallskip
2)~Order genes either randomly or use a predetermined fixed order. 
Defining a utility relation between genes, and ordering genes with respect to it can, in general, lead to problems that are considered intractable. Our results however suggest, it is important to understand relations between different genes, since genes are able to complement each other; this can lead to their increased mutual strength. On the other hand, random ordering adds to robustness of the underlying AIS. For an attacker, it is namely more difficult to deceive, since he does not know how genes are being used. It is currently an open question, how to impose a balanced solution.

\smallskip
3)~Genes cannot be considered in isolation. Our experiments show, when a detector matched an antigen under the $r$-contiguous matching rule, usually this match spanned over several genes. This motivates design of matching rules that would not limit matching to a few neighboring genes, offer more flexibility but still require that a gene remains a partly atomic unit.

\subsection{Learning and Detection}

Learning and detection is done by applying the mechanisms shown in Figs.~\ref{fig:det_gen} and~\ref{fig:non_self_det}. The detection itself is very straightforward. In the learning phase, a misbehavior-free period (see~\cite{aickelin2003dtl} on possibilities for circumventing this problem) is necessary so that nodes get a chance to learn what is the normal behavior. When implementing the learning phase, the designer gets to choose from two possibilities:
 
\smallskip
1)~Learning and detection at a node get implemented for each neighboring node separately. This means that different antigens have to get computed for each neighboring node, detector computation is different for each neighboring node and, subsequently, detection is different for each neighboring node. The advantage of this approach is that the node is able to directly determine which neighboring node misbehaves; the disadvantage is that $m$ instances ($m$ is the number of neighbors or node degree) of the negative selection mechanism have to get executed; this can be computationally prohibitive for sensor networks as $m$ can, in general, be equal to the total number of sensor. This allows for an easy Sybil attack~\cite{drozda2006ais} in which a neighbor would create several identities; the node would then be unable to recognize that these identities belong to the same neighbor. This approach was used in~\cite{sarafijanovic2004ais,leboudec2004ais}.

\smallskip
2)~Learning and detection at a node get implemented in a single instance for all neighboring nodes. This means a node is able to recognize anomaly (misbehavior) but it may be unable to determine which one from the $m$ neighboring nodes misbehaves. This implies that nodes would have to cooperate when detecting a misbehaving node, exchange anomaly information and be able to draw a conclusion from the obtained information. An argument for this approach is that in order to detect nodes that misbehave in collusion, it might be necessary to rely to some extent on information exchange among nodes, thus making this a natural solution to the problem. We have used this approach; a post-processing phase (using the list of misbehaving nodes) was necessary to determine whether a node was correctly flagged as misbehaving or not. 

\begin{figure}[!!!ht]
\begin{center}
  {
    \epsfig{file=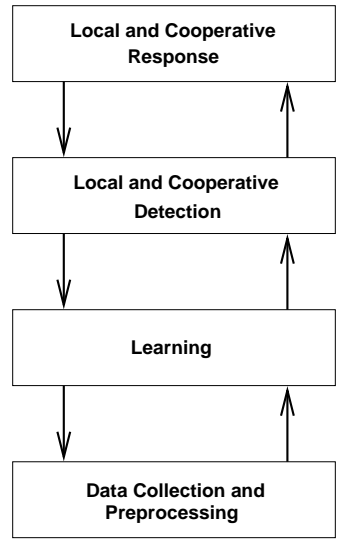, width=.55\linewidth}
  }
\caption{An four-layer architecture aimed at protecting sensor networks against misbehavior and abuse. }
\label{fig:arch}
\end{center}
\end{figure}

\smallskip
We find the second approach to be more suited for wireless sensor networks. It is namely less computationally demanding. We are unable, at this time, to estimate the frequency of a complete detector set computation.


Both approaches can be classified within the four-layer architecture (Fig.~\ref{fig:arch}) that we introduced in~\cite{drozda2005aah}. The lowermost layer, Data collection and preprocessing, corresponds to genes' computation and antigen construction. The Learning layer corresponds to the negative selection process. The next layer, Local and co-operative detection, suggests, an AIS should benefit from both local and cooperative detection. Both our setup and the setup described in~\cite{sarafijanovic2004ais,leboudec2004ais} only apply {\em local} detection. The uppermost layer, Local and co-operative response, implies, an AIS should also have the capability to undertake an action against one or several misbehaving nodes; this should be understood in a wider context of co-operating wireless devices acting in collusion in order to suppress or minimize the adverse impact of such misbehavior. To our best knowledge, there is currently no AIS implementation for sensor networks taking advantage of this layer.


\smallskip
\noindent
{\em Which $r$ is the correct one?}

An interesting technical problem is to tune the $r$ parameter for the $r$-contiguous matching rule so that the underlying AIS offers good detection and false positives rates. One possibility is a lengthy simulation study such as this one.  
Through multiparameter simulation we were to able to show that $r = 10$ offers the best performance for our setup. In~\cite{drozda06smd} we experimented with the idea of ``growing'' and ``shrinking'' detectors; this idea was motivated by~\cite{ji2004rvn}. The initial $r_0$ for a growing detector can be chosen as $r_0 = \lceil l/2 \rceil$, where $l$ is the detector length. The goal is to find the smallest $r$ such that a candidate detector does not match any self antigen. This means,  
initially, a larger (more specific) $r$ is chosen; the smallest $r$ that fulfills the above condition can be found through binary search. For shrinking detectors, the approach is reciprocal. Our goal was to show that such growing or shrinking detectors would offer a better detection or false positives rate. Short of proving this in a statistically significant manner, we observed that the growing detectors can be used for self tuning the $r$ parameter. The average $r$ value was close to the $r$ determined through simulation (the setup in that case was different from the one described in this document).

\subsection{Further Optimizations}

Our experiments show that only a small number of detectors get ever used (less than 5\%). The reason is, they get produced in a random way, not considering structure of the protocols. For example, a detector that is able to detect whether i) data packets got correctly transmitted and ii) 100\% of all MAC layers handshakes were incomplete is superfluous as this case should never happen. In~\cite{cayzer2005hgl}, the authors conclude: {\em ``... uniform coverage of non-self space is not only unnecessary, it is impractical; non-self space is too  big''}. Application driven knowledge can be used to set up a rule based system that would exclude infeasible detectors; see~\cite{dasgupta2002ibt} for a rule based system aimed at improved coverage of the non-self set. In~\cite{hofmeyr1999ida}, it is suggested that unused detectors should get deleted and the lifetime of useful detectors should be extended. 

\subsection{Misbehavior}

In a companion paper~\cite{drozda2006ais}, we have reviewed different types of misbehavior at the MAC, network and transport layers of the OSI protocol stack. We note that solutions to many of these attacks have been already proposed; these are however specific to a given attack. Additionally, due to the limitations of sensor networks, these solutions cannot be directly transfered.

The appeal of AIS based misbehavior detection rests on its simplicity and applicability in an environment that is extremely computationally and bandwidth limited. Misbehavior in sensor networks does not have to be executed by sensors themselves; one or several computationally more powerful platforms (laptops) can be used for the attack. On the other hand, a protection using such more advanced computational platforms is, due to e.g. the need to supply them continuously with electric power, harder to imagine. It would also create a point of special interest for the possible attackers.

\section{Experimental Setup}


The purpose of our experiments was to show that AIS are a
viable approach for detecting misbehavior in sensor networks. Furthermore, we wanted to cast light on internal performance of an AIS designed to protect sensor networks. One of our central goals was to provide an in-depth analysis of relative usefulness of genes.

{\em Definitions of input and output parameters:} The input parameters for our experiments were: $r$ parameter for the $r$-contiguous matching rule, the (desired) number of  detectors
 and misbehavior level. Misbehavior was modeled as random packet dropping at selected nodes.\\
\indent
The performance (output) measures were arithmetic averages and 95\% confidence intervals $ci_{95\%}$
of detection rate, number of false positives, real time to compute detectors, data traffic rate at nodes, 
number of iterations to compute detectors (number of random tries), number of non-valid detectors, 
number of different (unique) antigens in a run or a time window, and number of matches for each gene. The detection rate $dr$ is defined as $\frac{dns}{ns}$, where $dns$ is the number of detected non-self strings and $ns$ is the total number of non-self strings. A false positive in our definition is a string that is not self but can still be a result of anomaly that is identical with the effects of a misbehavior. A non-valid detector is a candidate detector that matches a self string and must therefore be removed. 

The number of matches for each gene was evaluated using the $r$-contiguous matching rule; we considered two cases: i) two bit-strings get matched from the left to the right and the first such a match will get reported (matching gets interrupted), ii) two bit-strings get matched from the left to the right and all possible matches will get reported. The time complexity of these two approaches is $O(r(l-r))$ and $\Theta(r(l-r))$, respectivelly; $r \leq l$, where $l$ is the bitstring length. The first approach is exactly what we used when computing the real time necessary for negative selection, the second approach was used when our goal was to evaluate relative usefulness of each gene.

\begin{figure*}[!!!tp]
\begin{center}

\noindent \fbox{

\begin{minipage}{15.5cm}
\small{    


(i)~Negative selection algorithm: random-generate-and-test. Implemented in C++, compiled with GNU g++ v4.0 with -O3 option.

(ii)~{\bf Input parameters:} 1. r-contiguous matching rule with $r=\{7, 10, 13, 16, 19, 22\}$. 2. Encoding: 5 genes each 10 bits long = 50 bits. 3. Number of detectors $\{500, 1000, 2000, 4000\}$. 4. Misbehavior level $\{10, 30, 50\%\}$ 5. Window size 500 seconds; 28 complete windows over 4-hour simulation time.

(iii)~{\bf Performance measures:} 
real time to compute detectors,
number of iterations to compute detectors, 
detection rate, false positives rate,
rate of non-valid detectors, 
data traffic rate at nodes, number of different antigens in a run, number of matches for each gene; their arithmetic averages and 95\% confidence intervals (where applicable).

(iv)~\textbf{Network topology:} Snapshot of movement modeled by random waypoint mobility model i.e. it is a static network. There were 1,718 nodes. The area was a square of 2,900m$\times$2,950m. The transmission range of transceivers was 100 meters.

(v)~\textbf{Number of connections:} 10 CBR (constant bit rate) connections.
\textbf{MAC protocol}: IEEE 802.11b DCF. \textbf{Routing protocol}: DSR. 
Other parameters: (i) Propagation path-loss
  model: two ray (ii) Channel frequency:
  2.4 GHz (iii) Topography: Line-of-sight (iv) Radio type: Accnoise (v)
  Network protocol: IPv4 (vi) Connection type: UDP. 

(vi)~\textbf{Injection rate:} 1 packet/second. 14,400 packets per connection were injected. Packet size was 512 bytes.

(vii)~The number of independent simulation runs for each combination of input parameters was 20. The simulation time was 4 hours. 

(viii)~ \textbf{Simulator used:} GlomoSim 2.03; hardware used: 30$\times$ Linux (SuSE 10.0) PC with 2GB RAM and Pentium 4 3GHz microprocessor.
}
\end{minipage}
}
\end{center}
\caption{Parameters used in the experiment.}
\label{fig:exp-parameters2}
\end{figure*}

{\bf Scenario description:}
We wanted to capture ``self'' and ``non-self'' packet traffic in a large enough synthetic static sensor network and test whether using an AIS we are able to recognize non-self, i.e. misbehavior. 

\begin{figure*}[!!!tp]
\begin{center}
  {
    \centerline{
    \subfigure[]{  
    \epsfig{file=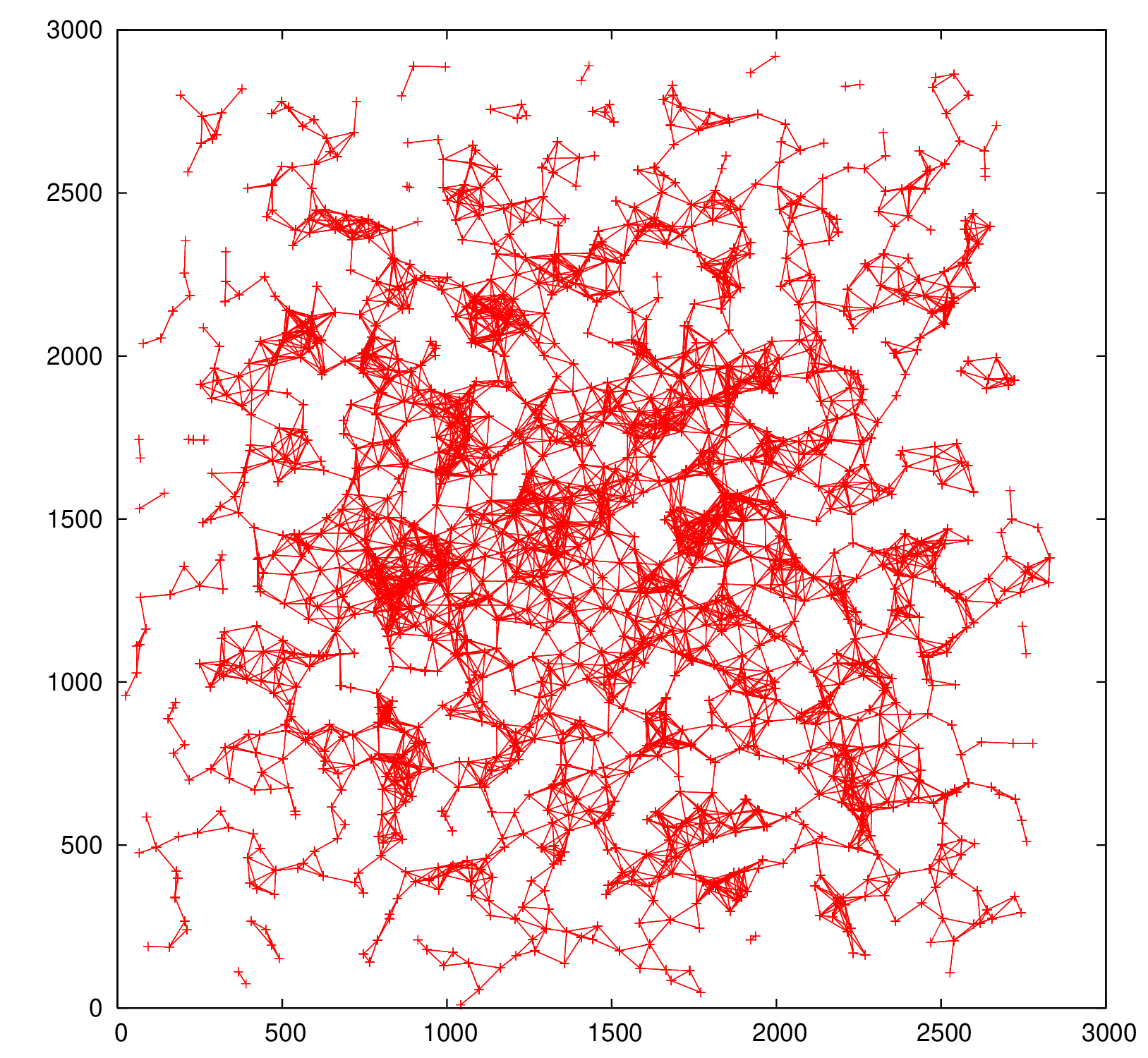, width=.4\linewidth}}
    \subfigure[]{
    \epsfig{file=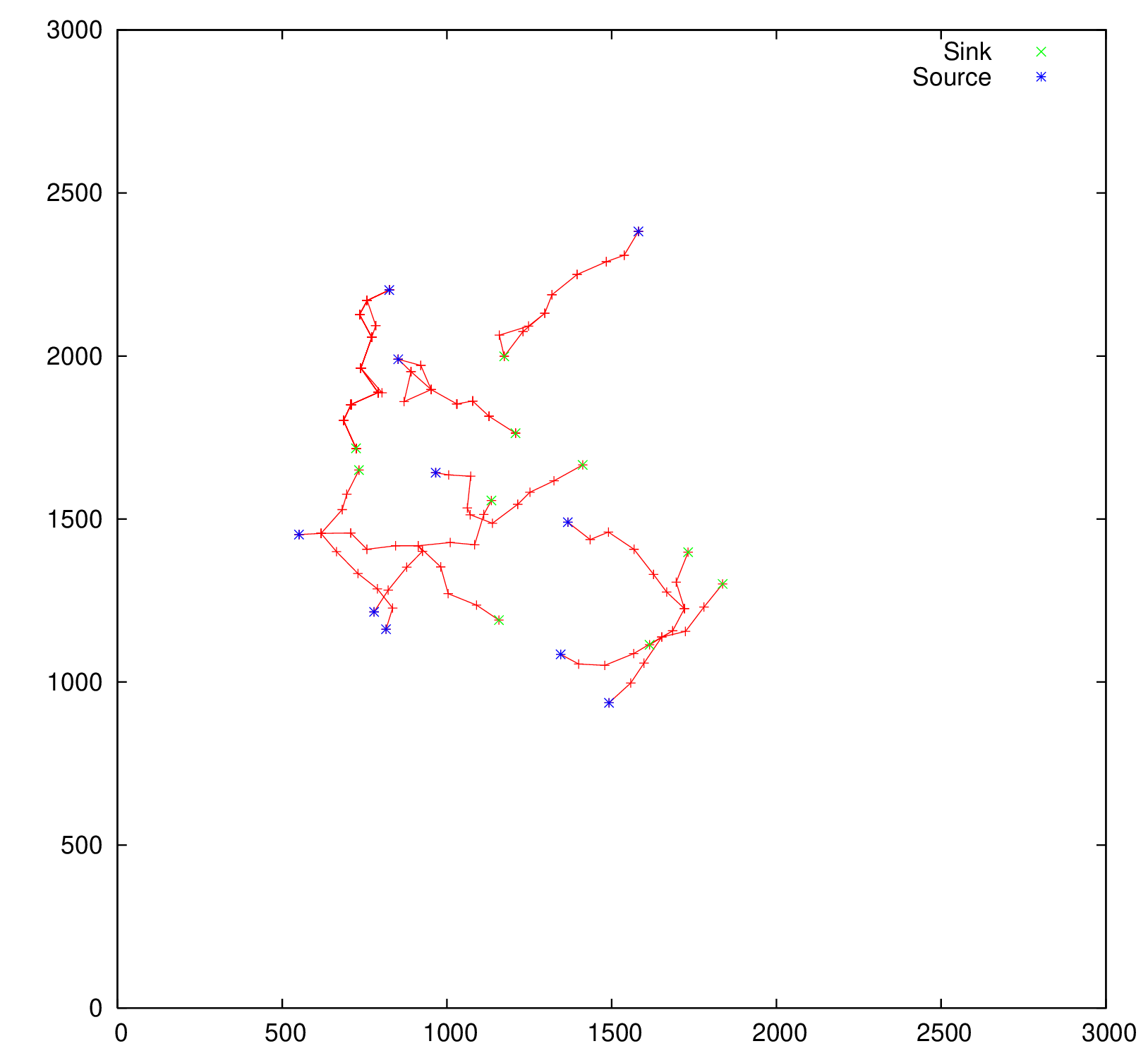, width=.4\linewidth}}
    }
  }
\caption{(a) Topology of our 1,718-node network with 100m radio radius. (b) Measured forwarding path of the 10 connections for a single simulation run without misbehavior; connections shown with all alternative forwarding routes, if they exist.}
\label{fig:topo}
\end{center}
\end{figure*}

The topology of this network was determined by making a {\em snapshot} of 1,718 mobile nodes (each with 100m radio radius) moving in a square area of 2,900m$\times$2,950m as prescribed by the random waypoint mobility model; see Figure~\ref{fig:topo}(a).  
The motivation in using this movement model and then creating a snapshot are the results in our previous paper~\cite{barrett2005upp} that deals with structural robustness of sensor network. Our preference was to use a slightly bigger network than it might be necessary, rather than using a network with unknown properties. The computational overhead is negligible; simulation real time mainly depends on the number of events that require processing. Idle nodes increase memory requirements, but memory availability at computers was in our case not a bottleneck.

We chose source and destination pairs for each connection so that several alternative independent routes exist; the idea was to benefit from route repair and route acquisition mechanisms of the DSR routing protocol, so that the added value of AIS based misbehavior detection is obvious.   

We used 10 CBR (Constant bit rate) connections.
 The connections were chosen so that their length is $\sim$7 hops and so that these connections share some common intermediate nodes; 
 see Figure~\ref{fig:topo}(b). 
For each packet received or sent by a node we have captured the following information: IP header type (UDP, 802.11 or DSR in this case), MAC frame type (RTS, CTS, DATA, ACK in the case of 802.11), current simulation clock, node address, next hop destination address, data packet source and destination address and packet size. 
 
{\em Encoding of self and non-self antigens:
} 
Each of the five genes was transformed in a 10-bit signature where each bit defines an interval\footnote{The interval encoding of genes is adapted from \cite{sarafijanovic2004ais}. This way only one of the 10 bits is set to 1, i.e. there are only 10 possible value levels that it is possible to encode in this case.} of a gene specific value range. We created self and non-self antigen strings by concatenation of the defined genes. Each self and non-self antigen has therefore a size of 50 bits. The interval representation was chosen in order to avoid carry-bits (the Gray coding is an alternative solution).

{\em Constructing the self and non-self sets:}
We have randomly chosen 28 non-overlapping 500-second windows in our 4-hour simulation. In each 500-second window self and non-self antigens are computed for each node. This was repeated 20 times for independent Glomosim runs. 

{\em Misbehavior modeling:}
Misbehavior is modeled as random data packet dropping (implemented at the network layer); data packets include both data packets from the transport layer as well as routing protocol packets. that should get dropped will simply not be inserted into the IP queue); we have randomly chosen 236 nodes and these were forced to drop $\{10, 30, 50\%\}$ of data packets. However, there were only 3-10 nodes with misbehavior and with a {\em statistically significant} number of packets for forwarding in each simulation run; see constraint C2 in Section~\ref{sec:eval}.
 
{\em Detection:} A neighboring node gets flagged as misbehaving, if a detector from the detector set matches an antigen. Since we used a single learning phase, we had to complement this process with some routing information analysis. This allowed us to determine, which one from the neighboring nodes is actually the misbehaving one. In the future, we plan to rely on co-operative detection in order to replace such a post-analysis.

{\em Simulation phases:} The experiment was done in four phases.
\begin{enumerate}
\item 20 independent Glomosim runs were done for one of $\{10, 30, 50\%\}$ misbehavior levels and ``normal'' traffic. Normal means that no misbehavior took place.
\item  Self and non-self antigen computation (encoding).
\item The 20 ``normal'' traffic runs were used to compute detectors. Given the 28 windows and 20 runs, the sample size was 20$\times$28 = 560, i.e. detectors at each node were discriminated against 560 self antigens.
\item Using the runs with $\{10, 30, 50\%\}$ misbehavior levels,  the process shown in Fig.~\ref{fig:non_self_det} was used for detection; we restricted ourselves to nodes that had in both the normal and misbehavior traffic at least a certain number of data packets to forward (packet threshold). 
\end{enumerate}

The experiment was then repeated with different $r$, desired number of detectors and misbehavior level. 

The parameters for this experiment are summarized in Fig.~\ref{fig:exp-parameters2}. The injection rate and packet sizes were chosen in order to comply with usual data rates of sensors (e.g. 38.4kbps for Mica2; see~\cite{xbow}). We chose the Glomosim simulator~\cite{bajaj1999gsn} over other options (most notably ns2) because of its better scaling characteristics~\cite{barr2005jea} and our familiarity with the tool.


\begin{figure*}[!!!tp]
\begin{center}
    \subfigure[Real time to compute the desired number of detectors at a node; $ci_{95\%} < 1\%$.]{
    \epsfig{file=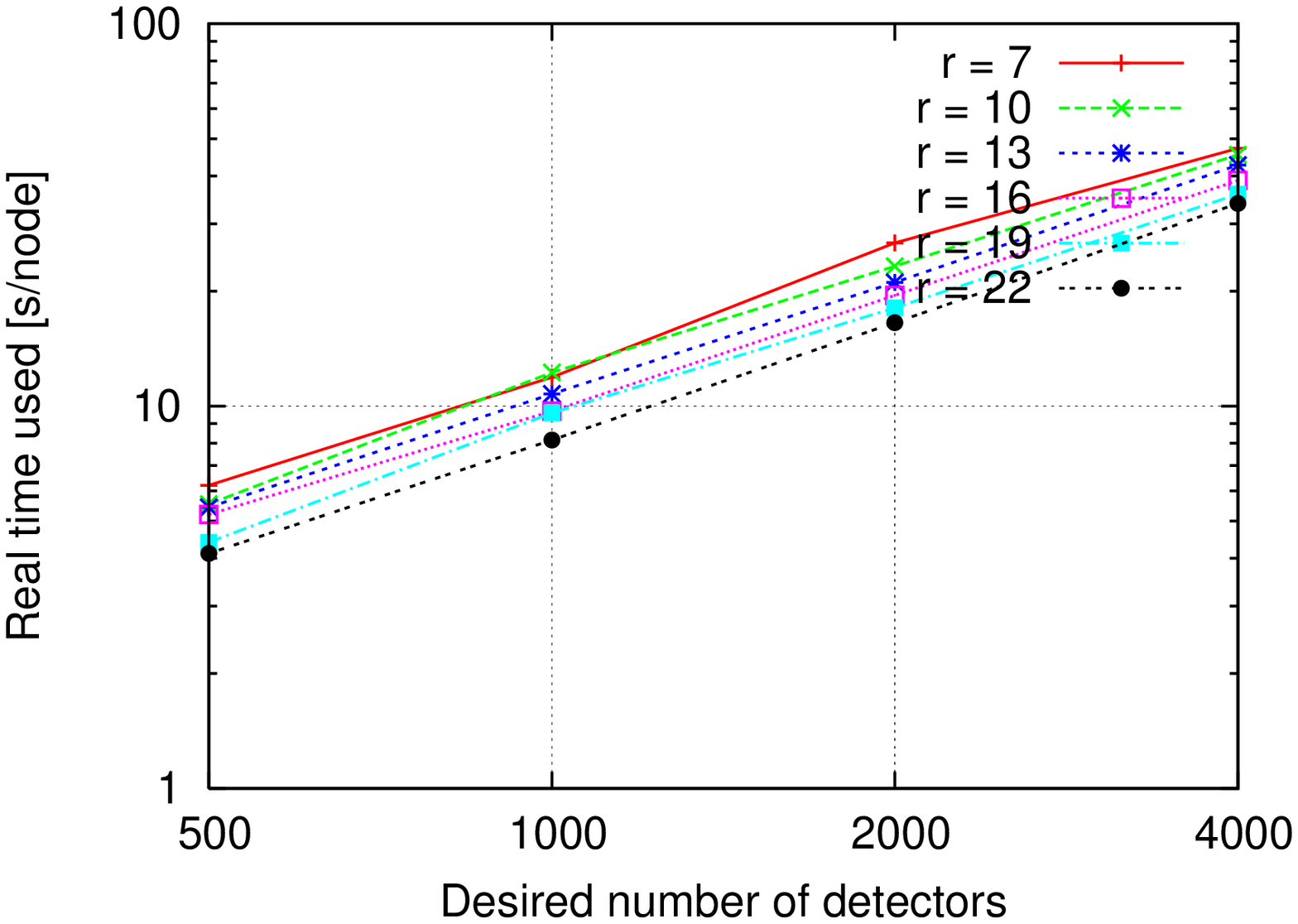, width=.32\linewidth}}
    \subfigure[Rate of non-valid detectors; for $r \leq 13$ is $ci_{95\%} < 1\%$, for $r \geq 16$ is the sample size not significant.]{
    \epsfig{file=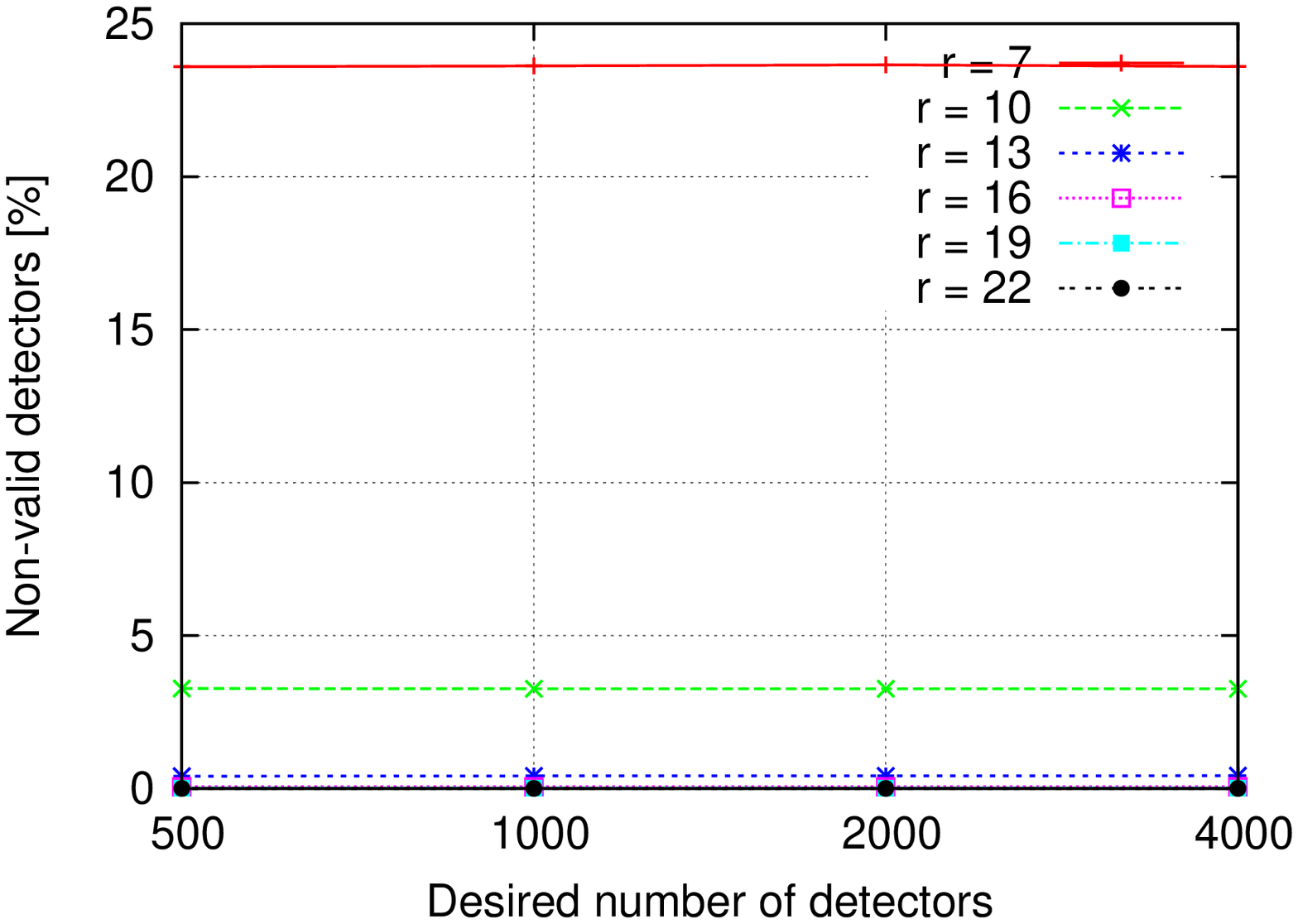, width=.32\linewidth}}
    \subfigure[Number of iterations needed in order to compute the desired number of detectors; for $r \geq 10$ is $ci_{95\%}< 1\%$, for $r=7$ is $ci_{95\%} < 2\%$.]{
    \epsfig{file=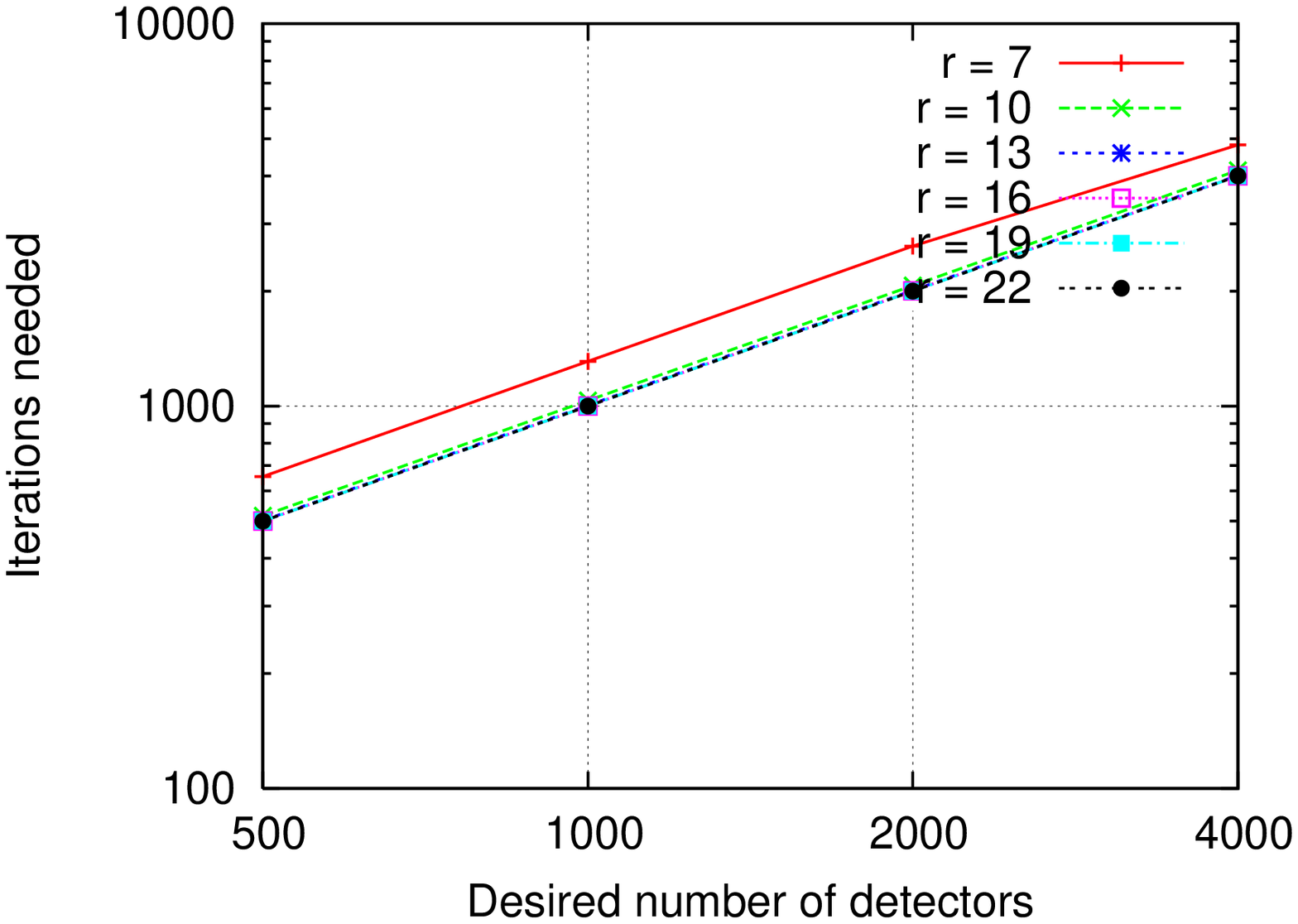, width=.32\linewidth}}  
\caption{Performance of detectors computation.}
\label{fig:300m}
\end{center}
\end{figure*}

\begin{figure*}[!!!tp]
\begin{center}
  {
    \subfigure[Detection rate vs packet threshold; conf. interval ranges: for mis. level $10\%$ is $ci_{95\%}$ = 3.8-19.8\%; for $30\%$ is $ci_{95\%}$ = 11.9-15.9\%; for $50\%$ is $ci_{95\%}$ = 11.0-14.2\%.]{
    \epsfig{file=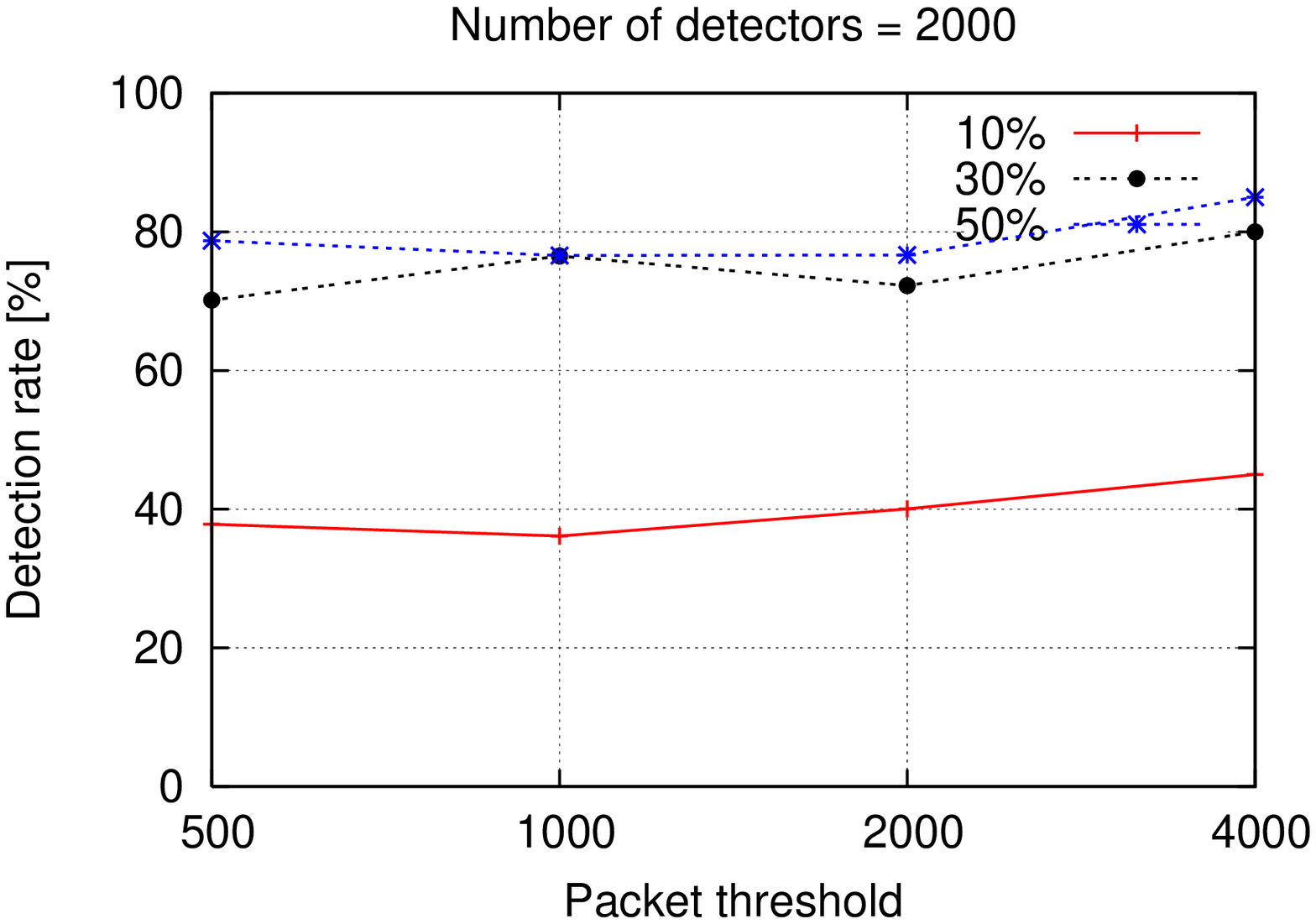, width=.32\linewidth}}
    \subfigure[Detection rate vs $r$; $ci_{95\%}$ range similar to (a).]{
    \epsfig{file=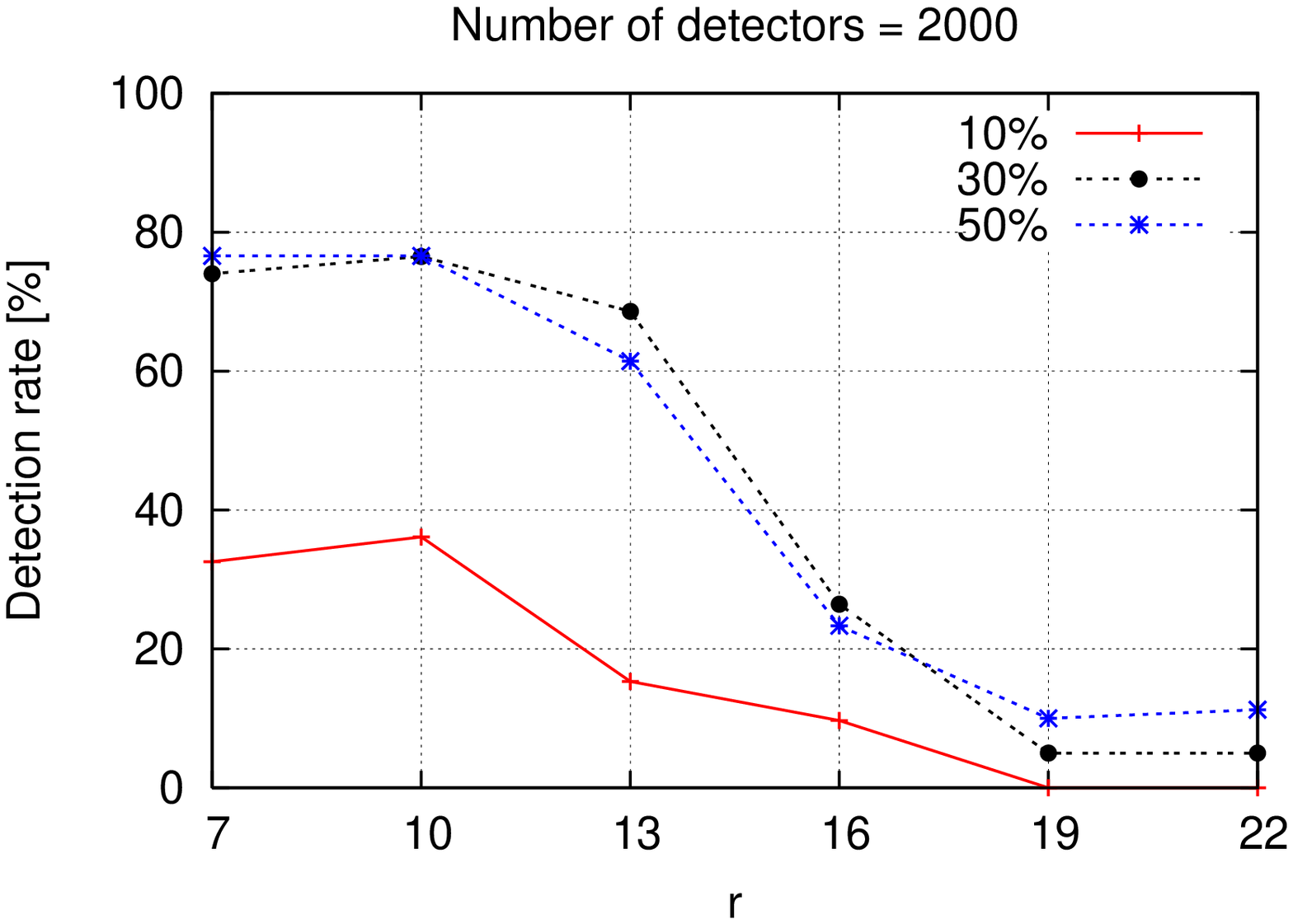, width=.32\linewidth}}
    \subfigure[Number of false positives; for $r \leq 10$ is $ci_{95\%}$ = 0.47-0.68, for $r \geq 13$ is the sample size not significant.]{
    \epsfig{file=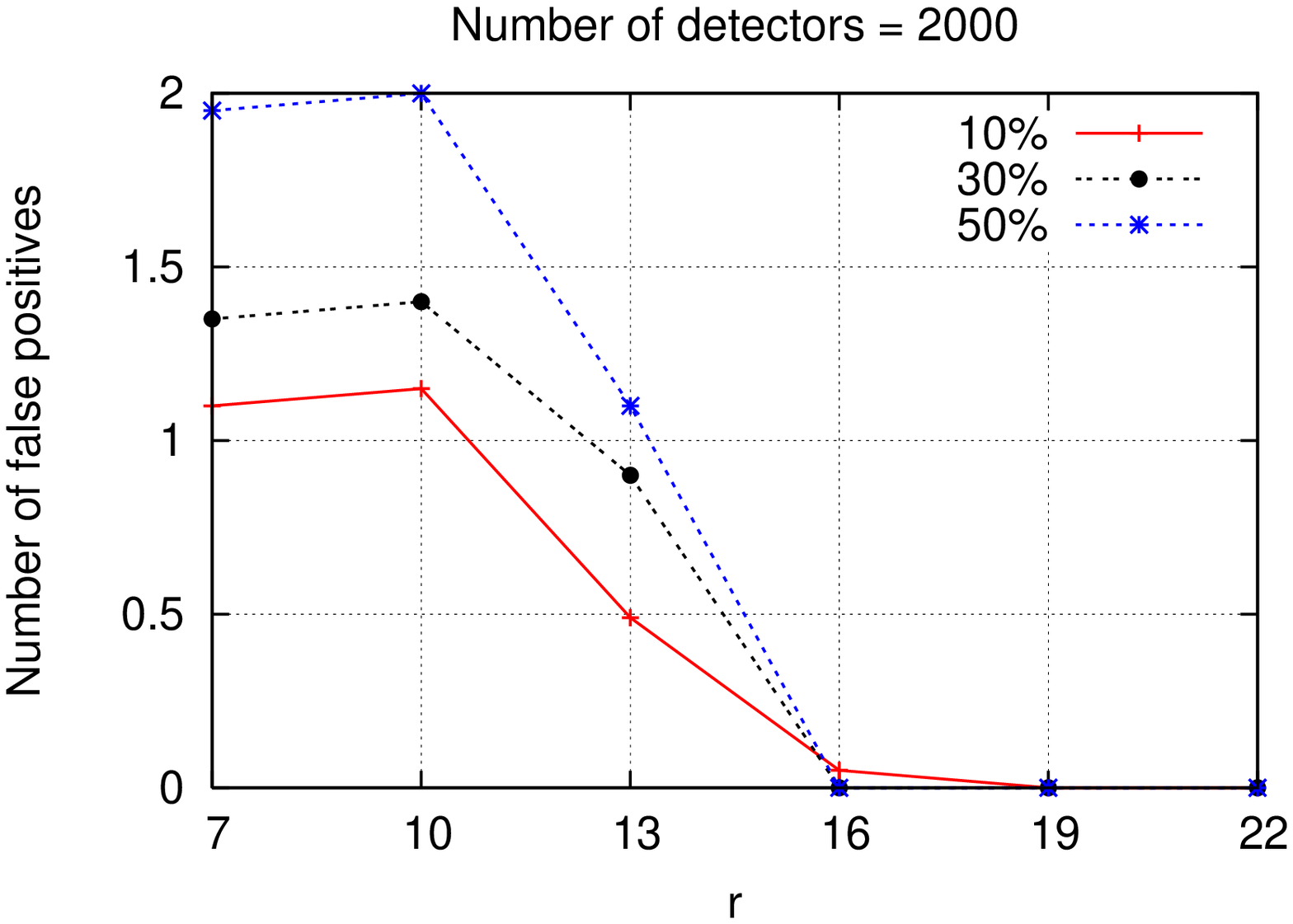, width=.32\linewidth}}
  }
\caption{Performance of misbehavior detection. Misbehavior level = \{10, 30, 50\}\%. In (a) $r = 10$, in (b) and (c) the packet threshold was 1000. }
\label{fig:perform}
\end{center}
\end{figure*}

\begin{figure*}[!!!tp]
\begin{center}
  {
    \subfigure[Total number of runs with window threshold $\geq 14$.]{
    \epsfig{file=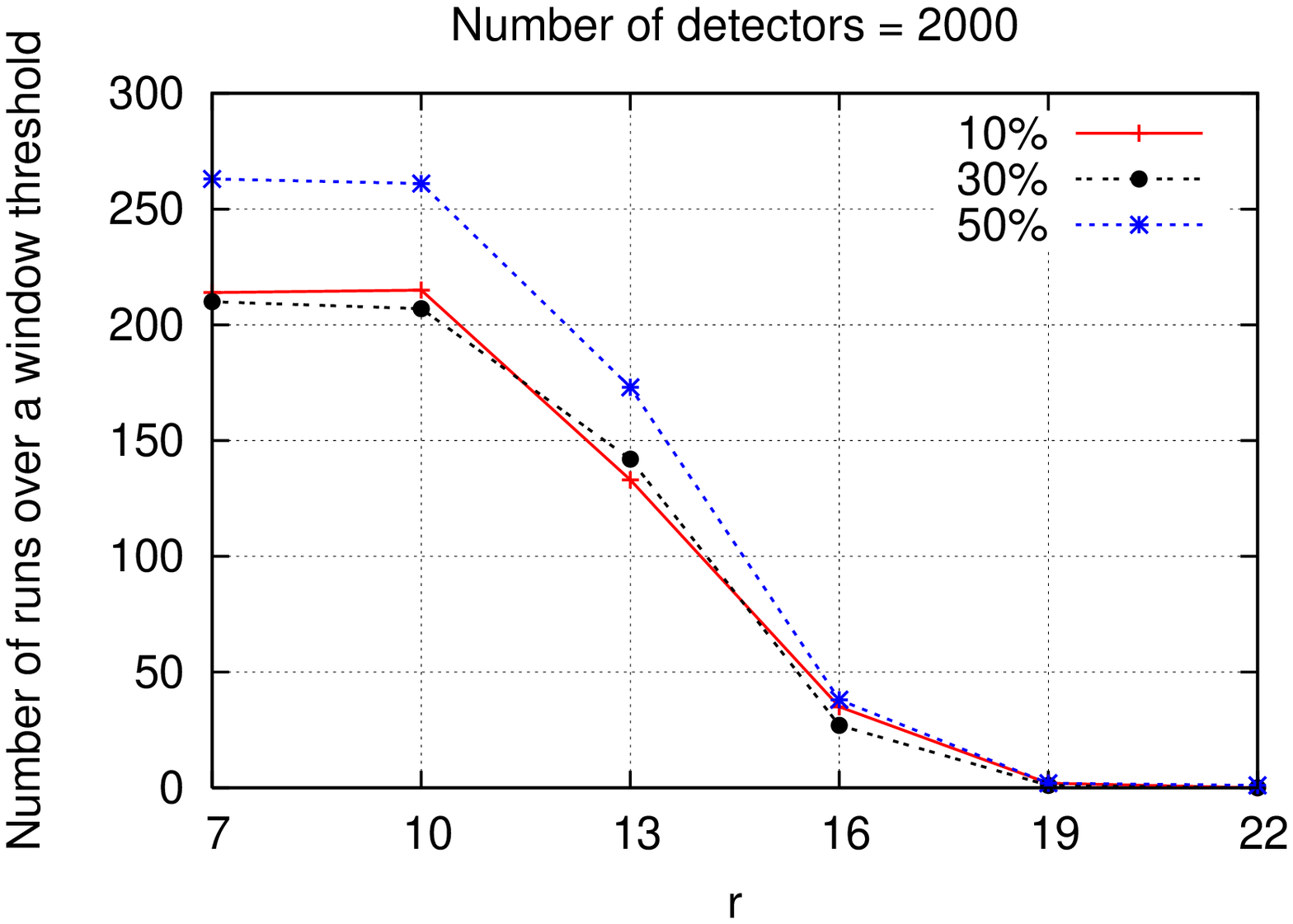, width=.32\linewidth}}
    \subfigure[The number of unique detectors that matched an antigen in a run. Conf. interval range for $7 \leq r \leq 13$ is $ci_{95\%}$ = 6.5-10.1\%.]{
    \epsfig{file=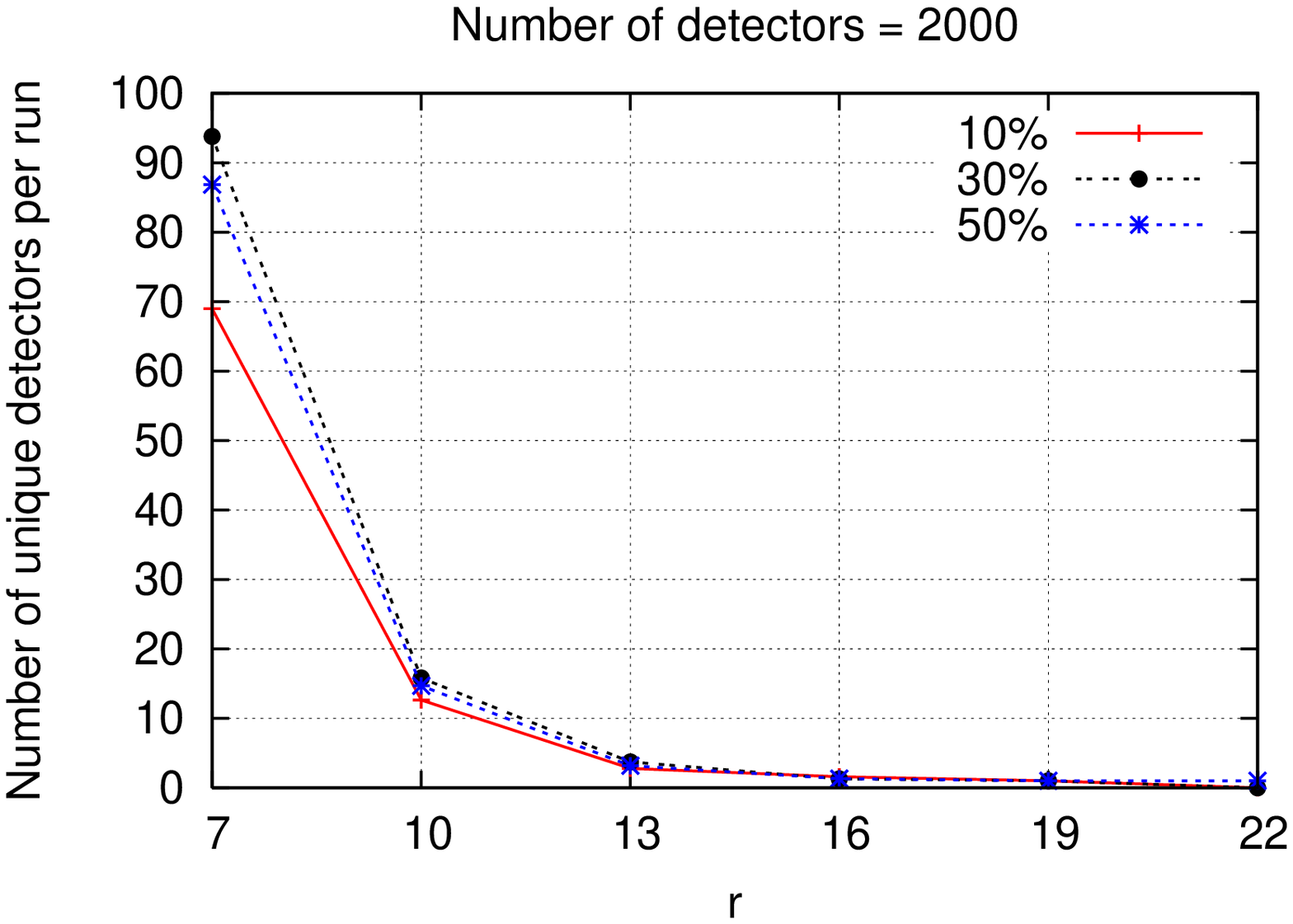, width=.32\linewidth}}
    \subfigure[The number of unique detectors that matched an antigen in a window; each run has 28 windows. Conf. interval range: $ci_{95\%} <$ 0.16\%.]{
    \epsfig{file=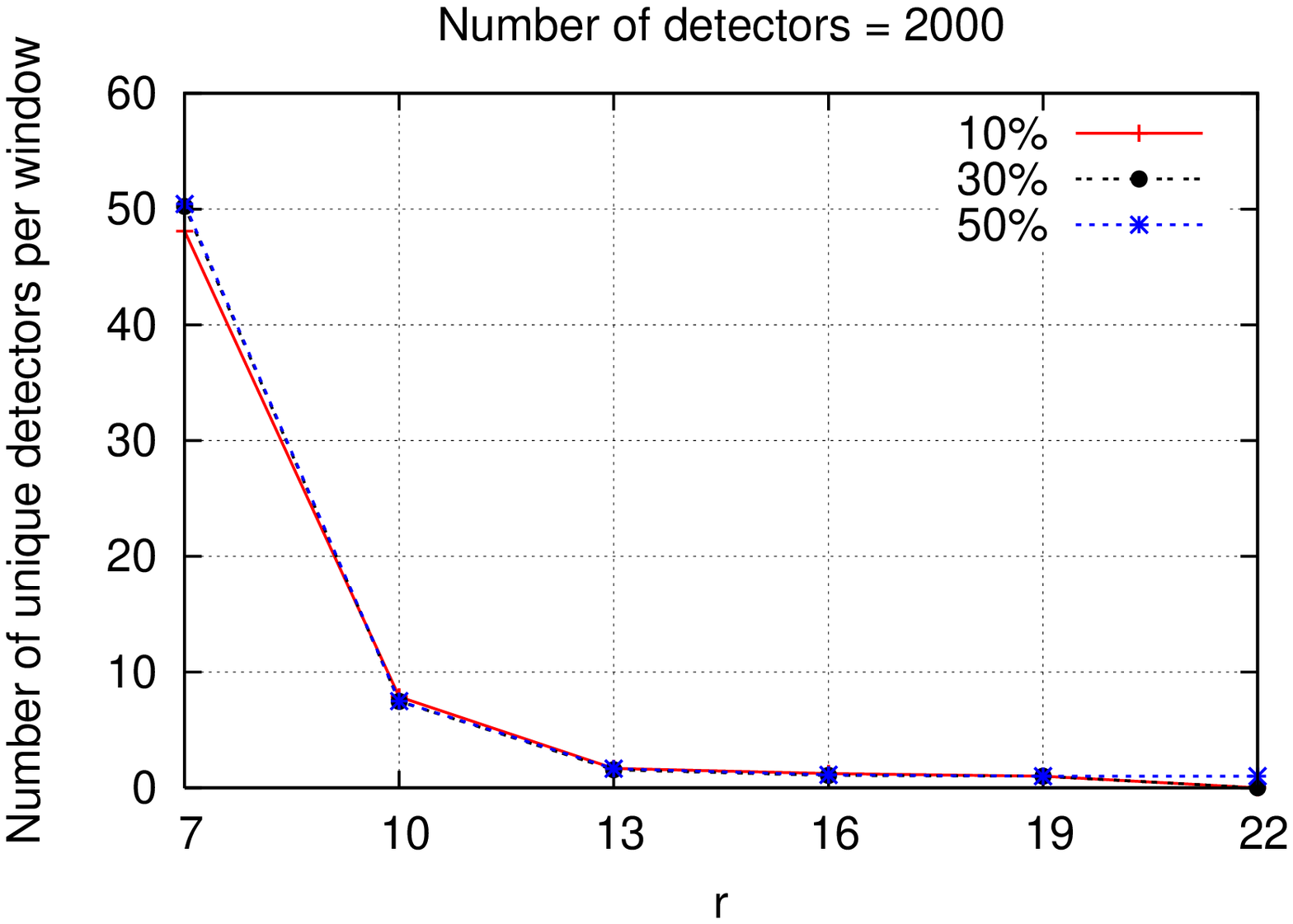, width=.32\linewidth}}
  }
\caption{Window threshold and detector related performance measures.}
\label{fig:det_used1}
\end{center}
\end{figure*}

\begin{figure*}[!!!tp]
\begin{center}
  {
    \subfigure[Number of unique antigens per run. Conf. interval range for $7 \leq r \leq 13$ is $ci_{95\%}$ = 5.3-8.9\%.]{
    \epsfig{file=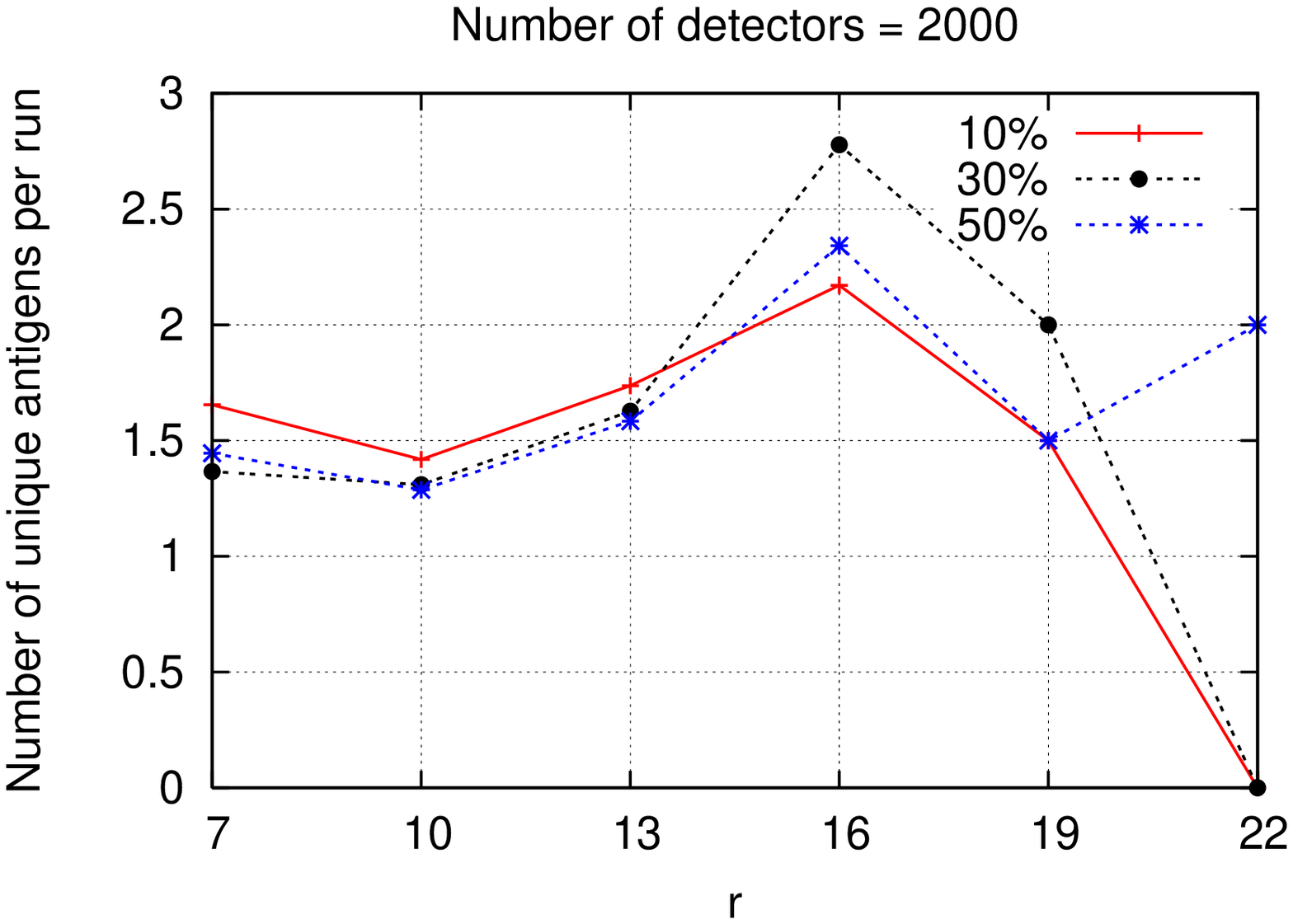, width=.32\linewidth}}
    \subfigure[Total number of matches; $single$ = one gene got matched, $multiple$ = more than one gene got matched.]{
    \epsfig{file=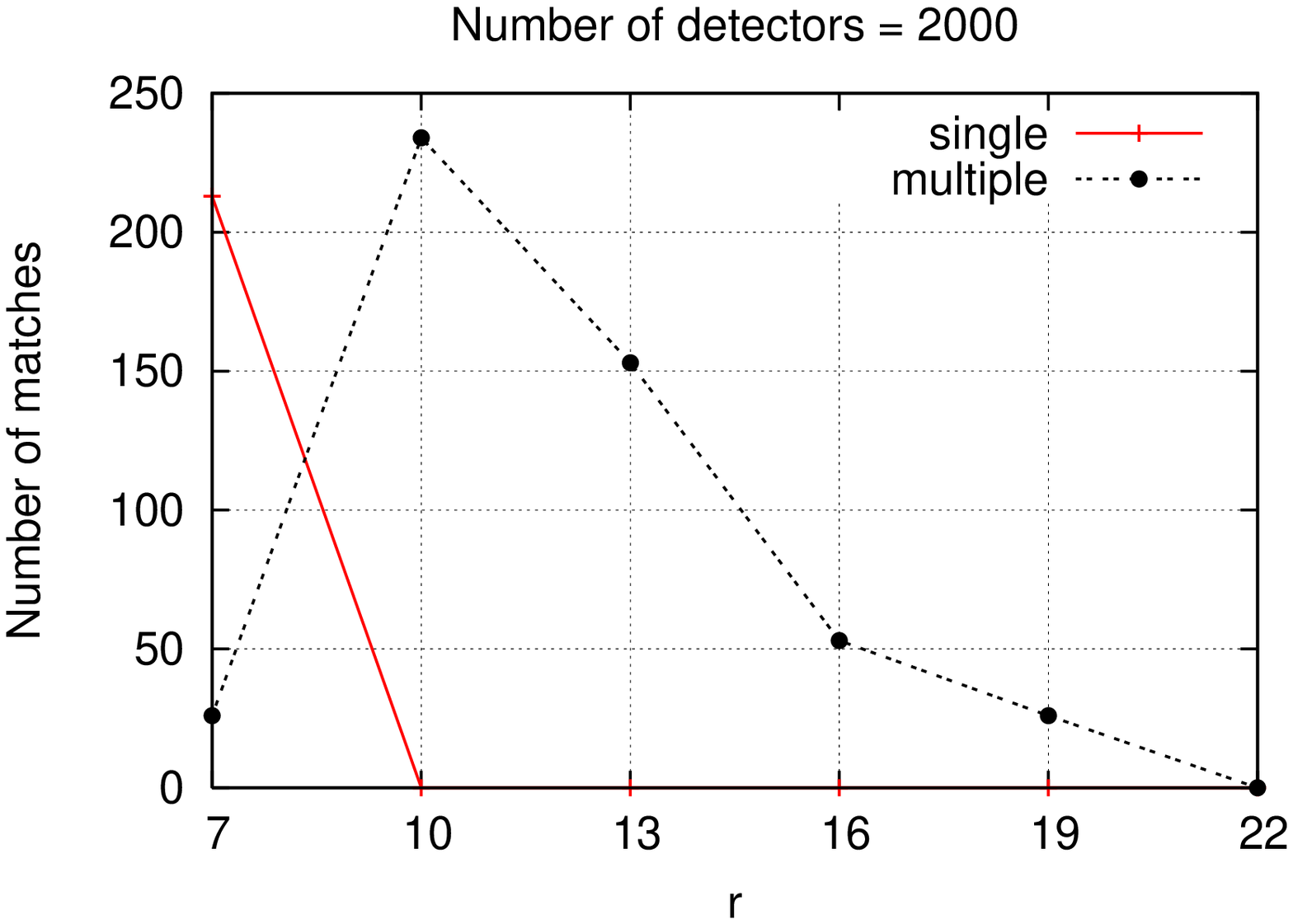, width=.32\linewidth}}
    \subfigure[Total number of matches for Gene \#1.]{
    \epsfig{file=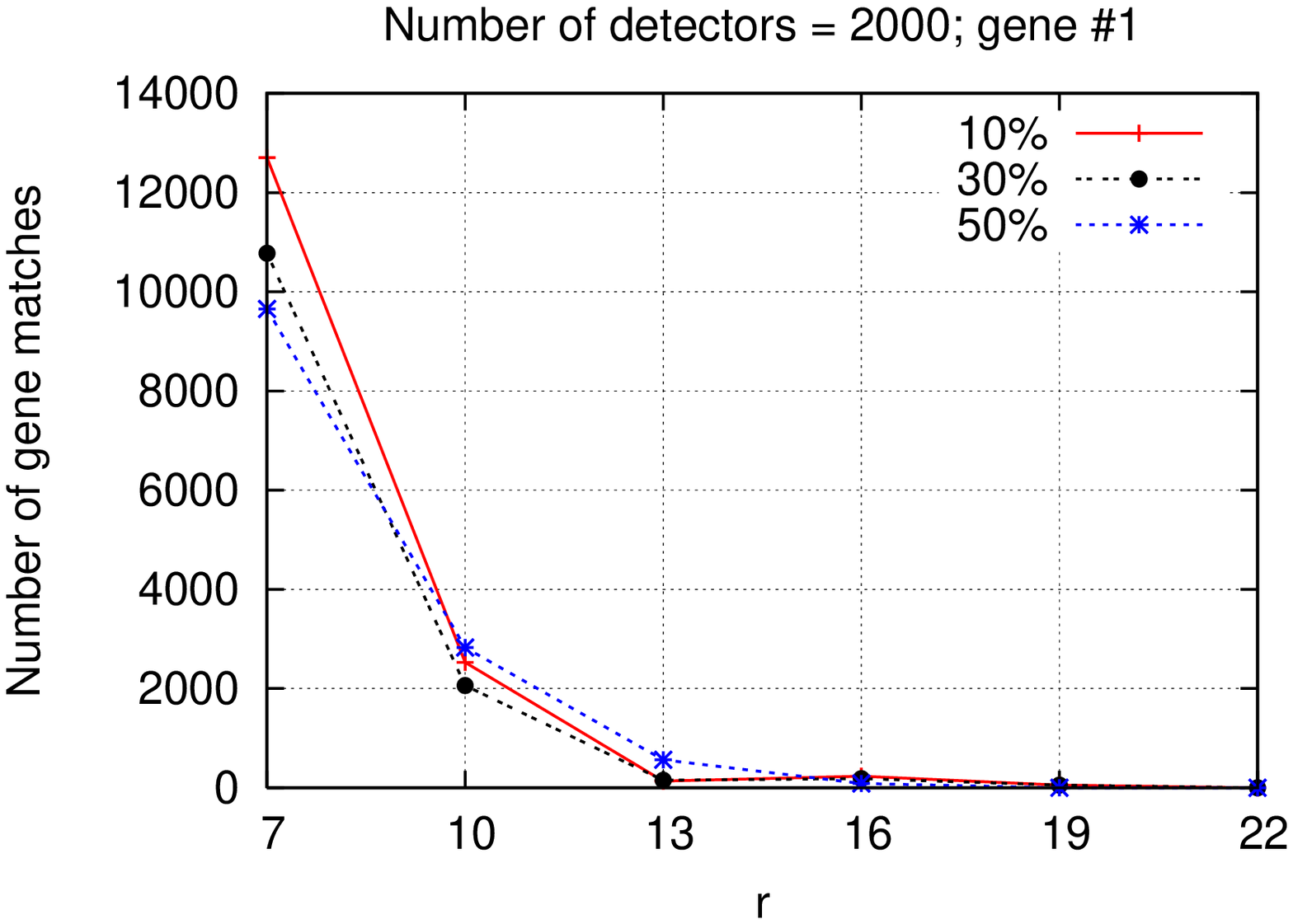, width=.32\linewidth}}
  }
\caption{Antigen and gene related performance measures.}
\label{fig:det_used2}
\end{center}
\end{figure*}

\begin{figure*}[!!!tp]
\begin{center}
  {
    \subfigure[]{
    \epsfig{file=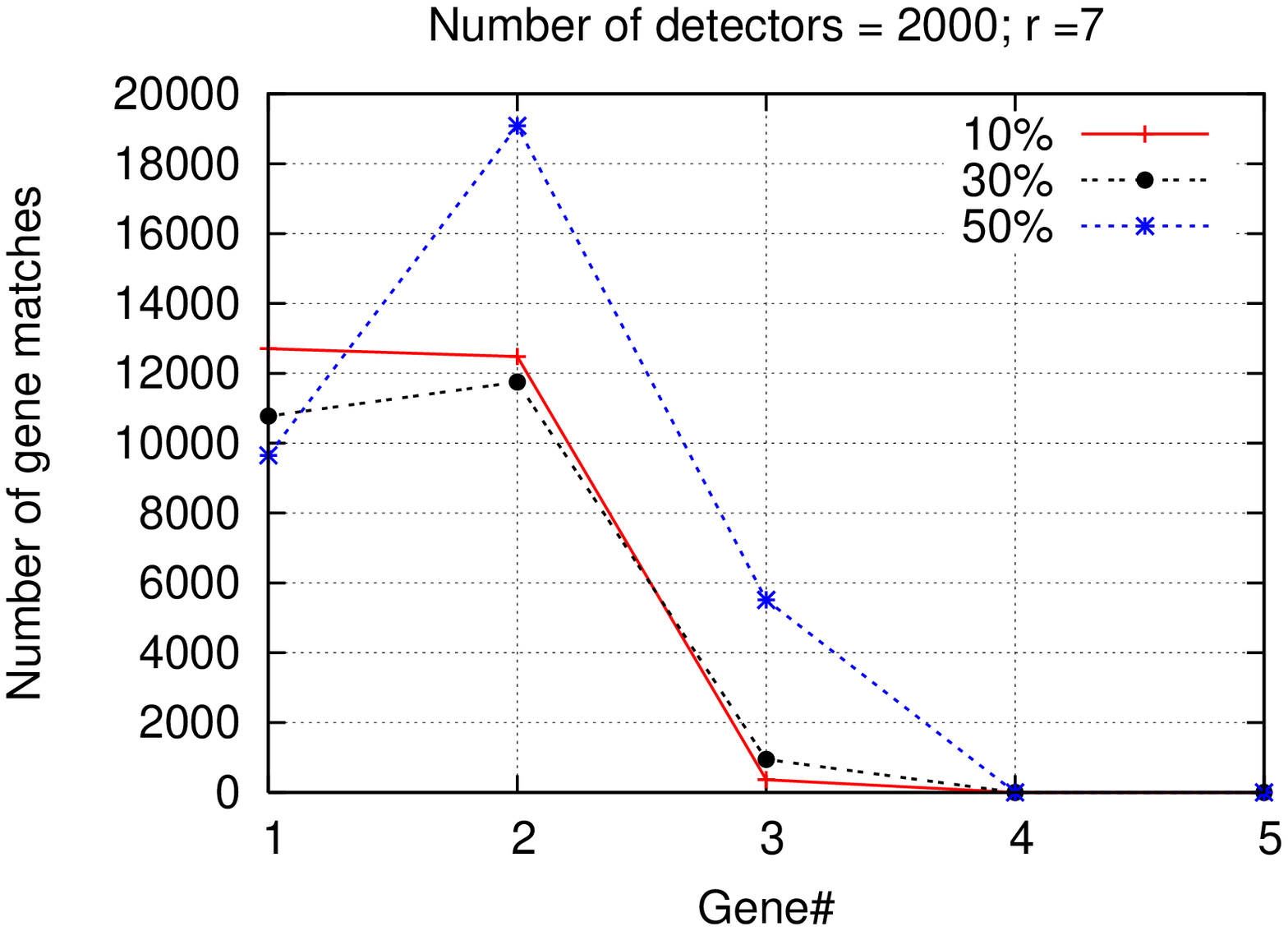, width=.32\linewidth}}
    \subfigure[]{
    \epsfig{file=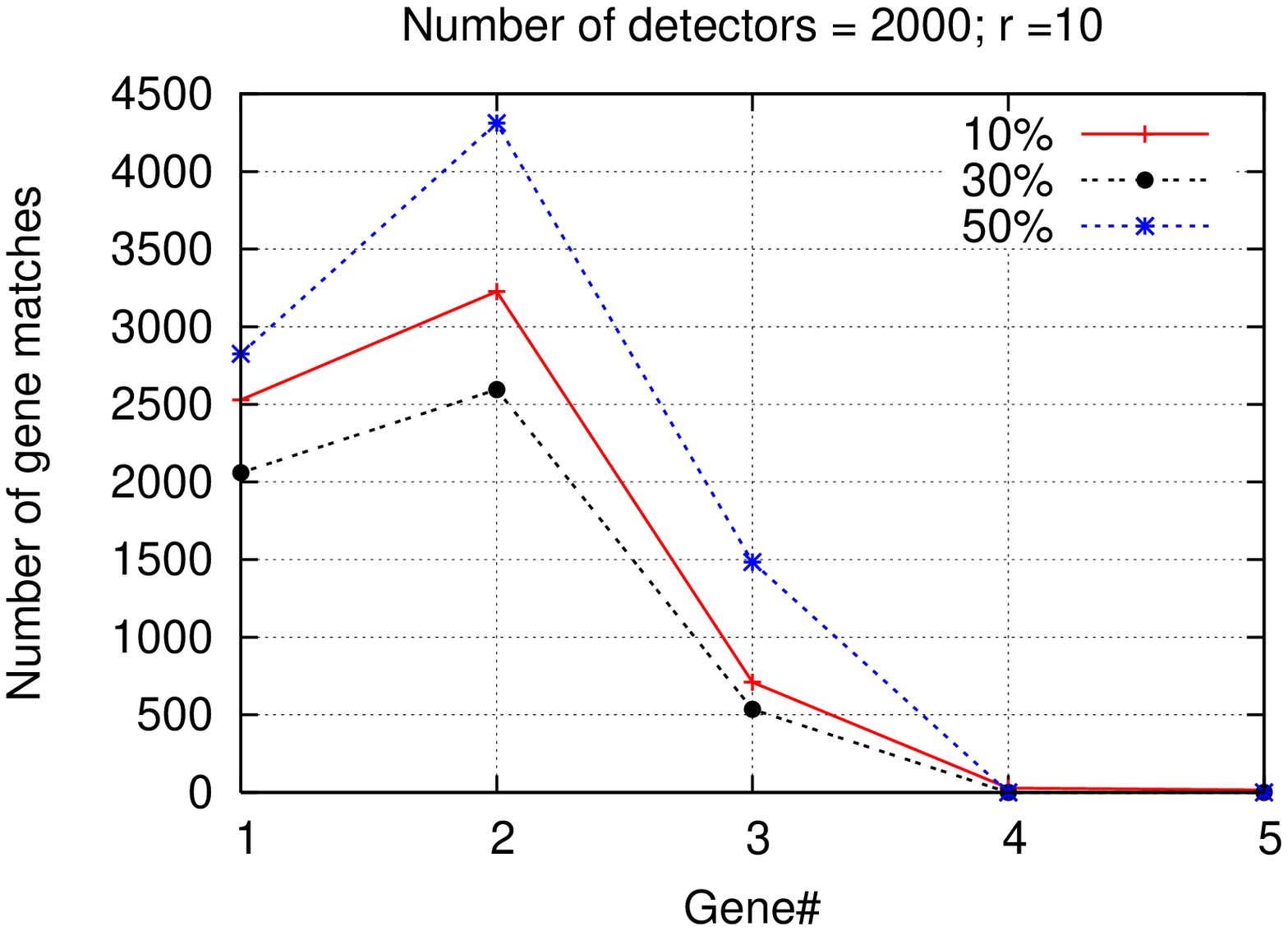, width=.32\linewidth}}
    \subfigure[]{
    \epsfig{file=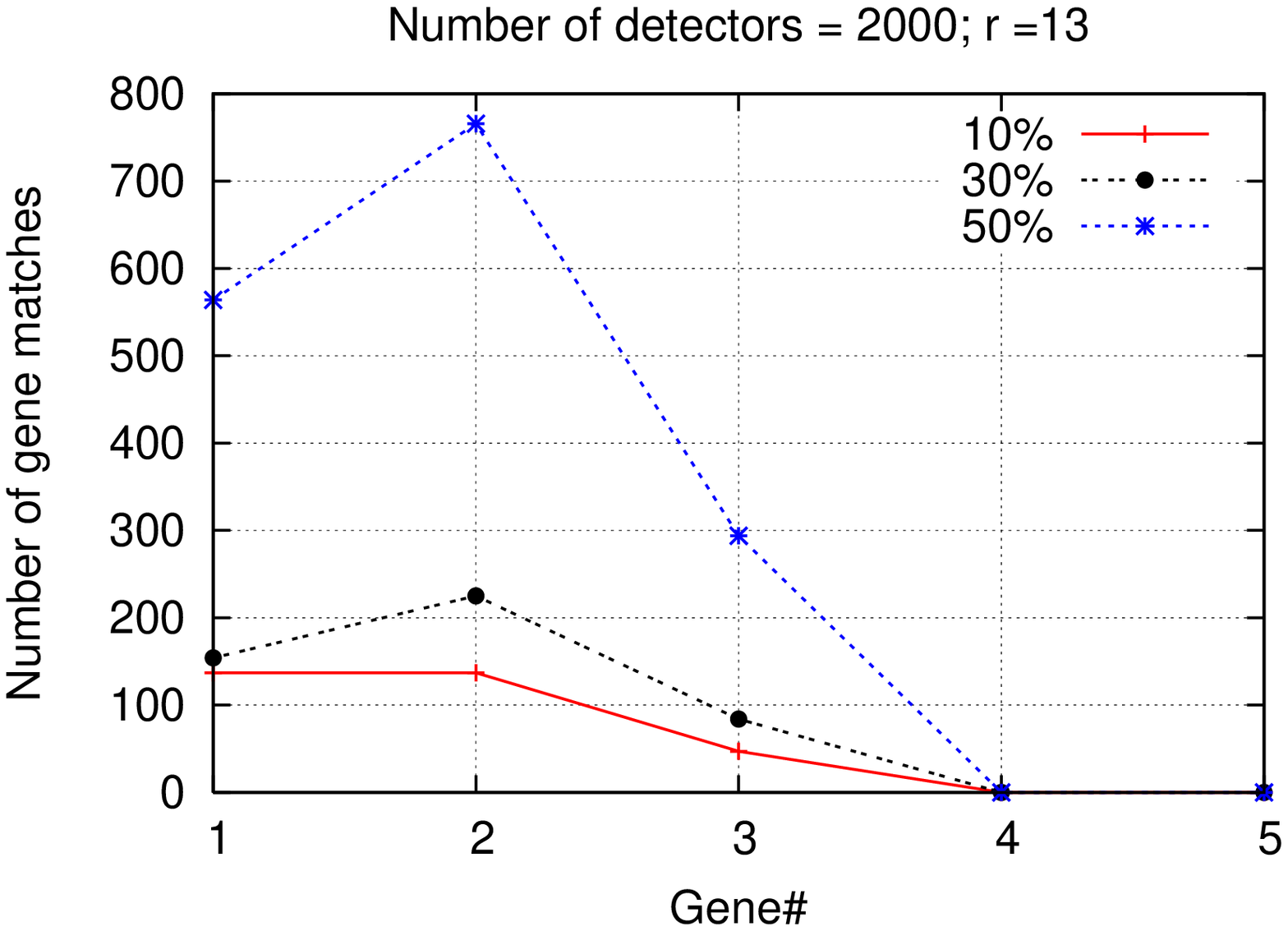, width=.32\linewidth}}
  }
\caption{Performance of Genes \#1 through \#5 for  the number of detectors = $2000$ and (a) $r = 7$, (b) $r = 10$, (c) $r = 13$. }
\label{fig:det_used3}
\end{center}
\end{figure*}

\section{Results Evaluation}
\label{sec:eval}

When evaluating our results we define two additional constraints:
\begin{enumerate}
\item[C1.] We define a node to be detected as misbehaving if it gets flagged in at least 14 out of the 28 possible windows. This notion indirectly defines the time until a node is pronounced to be misbehaving. We call this a {\em window threshold}. 

\item[C2.] A node $s_i$ has to forward in average at least $m$ packets over the 20 runs in both the ``normal'' and misbehavior cases in order to be included into our statistics. This constraint was set in order to make the detection process more reliable. It is dubious to flag a neighboring node of $s_i$ as misbehaving, if it is based on ``normal'' runs or runs with misbehavior, in which node $s_i$ had no data packets to forward (he was not on a routing path). We call this a {\em packet threshold}; $m$ was in our simulations chosen from $\{500, 1000, 2000, 4000\}$. {\em Example:} for a fixed set of input parameters, a node forwarded in the ``normal'' runs in average 1,250 packets and in the misbehavior runs (with e.g. level 30\%) 750 packets. The node $s_i$ would be considered for misbehavior detection if $m = 500$, but not if $m \geq 1000$. In other words, a node has to get a chance to learn what is ``normal'' and then to use this knowledge on a non-empty packet stream.
\end{enumerate}


\subsection{Overall Performance}

The results related to computation of detectors are shown in Figure~\ref{fig:300m}. In our experiments we have considered the desired number of detectors to be max. 4,000; over this threshold the computational requirements might be too high for current sensor devices. We remind the reader, each time the $r$ parameter is incremented by $1$, the number of detectors should double in order to make these two cases comparable. 

Figure~\ref{fig:300m}(a) shows the real time needed to compute the desired set of detectors. We can see the real time necessary increases proportionally with the desired number of detectors; this complies with the theoretical results presented in~\cite{dhaeseleer1996iac}.  Figure~\ref{fig:300m}(b) shows the percentage of non-valid detectors, i.e. candidate detectors that were found to match a self string (see Figure~\ref{fig:det_gen}). This result points to where the optimal operation point of an AIS might lie with respect to the choice of $r$ parameter and the choice of a fixed number of detectors to compute. We remind the reader, the larger is the $r$ parameter the smaller is the probability that a detector will match a self string. Therefore overhead connected to choosing the $r$ parameter prohibitively small should be considered when designing an AIS. Figure~\ref{fig:300m}(c) shows the total number of generate-and-test tries needed for computation of detector set of a fixed size; the 95\% confidence interval is less than $2\%$.

In Figure~\ref{fig:perform}(a) we show the dependence of detection ratio on the packet threshold. 
We conclude that except for some extremely low threshold values (not shown) the detection rate stays constant. This figure also shows that when misbehavior level was set very low, i.e. 10\%, the AIS struggled to detect misbehaving nodes. This is partly a result of our coarse encoding with only 10 different levels. 

At the 30 and 50\% misbehaving levels the detection rate stays solid at about 70-85\%. 
The range of the 95\% confidence interval of detection rate is 3.8-19.8\%. 
The fact that the detection rate did not get closer to 100\% suggests, either the implemented genes are not sufficient, detection should be extended to protocols at other layers of the OSI protocol stack, a different ordering of genes should have been applied or our ten level encoding was too coarse.
It also implicates that watchdog based genes (though they perfectly fit the implemented misbehavior) should not be used in isolation, and in general, that the choice of genes has to be very careful. 

Figure~\ref{fig:perform}(b) shows the impact of $r$ on detection rate. When $r=\{7, 10\}$ the AIS performs well, for $r > 10$ the detection rate decreases. This is caused by the inadequate numbers of detectors used at higher levels of $r$ (we limited ourselves to max. 4,000 detectors).

Figure~\ref{fig:perform}(c) shows the number of false positives. We remind that in our definition false positives are both nodes that do not drop any packets and nodes that drop packets due to other reasons than misbehavior.

In a separate experiment we studied whether the 4-hour (560 samples) simulation time was enough to capture the diversity of the self behavior. 
This was done by trying to detect misbehavior in 20 independent misbehavior-free  Glomosim runs (different from those used to compute detectors).
We report that we did not observe a single case of an autoimmune reaction.

\subsection{Detailed Performance}

In Fig.~\ref{fig:det_used1}(a) we show the total number of runs in which a node was identified as misbehaving. The steep decline for values $r>10$ (in this and other figures) documents that in these cases it was necessary to produce a higher number of detectors in order to cover the non-self antigen space.
The higher the $r$, the higher is the specificity of a detector, this means that it is able to match a smaller set of non-self antigens. 

In Fig.~\ref{fig:det_used1}(b) and (c) we show the number of detectors that got matched during the detection phase (see Fig.~\ref{fig:non_self_det}). Fig.~(b) shows the number of detectors matched per run, Fig.~(c) shows the number of detectors matched per window. Fig.~(b) is an upper estimate on the number of unique detectors needed in a single run. Given that the total number of detectors was 2,000, there were less than 5\% detectors that would get used in the detection phase. The tight confidence intervals\footnote{For practical reasons we show $ci_{95\%}$ only for $7 \leq r \leq 13$.} for the number of unique detectors matched per window (see Fig.~(c)) is a direct consequence of the small variability of antigens as shown in Fig.~\ref{fig:det_used2}(a). 

Fig.~\ref{fig:det_used2}(a) shows the number of unique antigens that were subject to classification into self or non-self. The average for $r = \{7,10\}$ is about 1.5. This fact does not directly imply that the variability of the data traffic would be inadequate. It is rather a direct consequence of our choice of genes and their encoding (we only used 10 value levels for encoding).
 Fig.~\ref{fig:det_used2}(b) shows the number of matches between a detector and an antigen in the following way. When a detector under the $r$-contiguous matching rule matches only a single gene within an antigen, we would increment the ``single'' counter. Otherwise, we would increment the ``multiple'' counter. It is obvious that with increasing $r$, it gets more and more probable that a detector would match more than a single gene. The interesting fact is that the detection rate for both $r=7$ and $r=10$ is about 80\% (see Fig.~\ref{fig:perform}(a)) and that the rate of non-valid detectors is very different (see Fig.~\ref{fig:300m}(b)). This means that an interaction between genes has positively affected the later performance measure, without sacrificing on the former one. This leads to a conlusion that genes should not be considered in isolation.


Fig.~\ref{fig:det_used2}(c) shows the performance of Gene \#1. The number of matches shows that this gene contributed to the overall detection performance of our AIS. Figs.~\ref{fig:det_used3}(a-c) sum up performance of the five genes for different values of $r$. Again, an interesting fact is the contribution of Gene \#1 to the overall detection performance. The usefulness of Gene \#2 was largely expected as this gene was tailored for the kind of misbehavior that we implemented. The other three genes came out as marginally useful. The importance of the somewhat surprising performance of Gene \#1 is that it can be computed in a simplistic way and does not require continuous operation of a node.

\subsection{The Impact of Data Traffic Pattern}

In an additional experiment, we examined the impact of data traffic pattern on the performance. We used two different data traffic models: the constant bit rate (CBR) and a Poisson distributed data traffic. In many scenarios, sensors are expected to take measurements in constant intervals and, subsequently, send them out for processing. This would create a constant bit rate traffic. Poisson distributed traffic could be a result of sensors taking measurements in an event-driven fashion. For example, a sensor would take a measurement only when a target object (e.g. a person) happens to be in its vicinity. 

The setup for this experiment was similar to that presented in Fig.~\ref{fig:exp-parameters2} with the additional fact that the data traffic model would now become an input parameter. With the goal to reduce complexity of the experimental setup, we fixed $r=10$ and we only considered cases with $500$ and $2000$ detectors. In order to match the CBR traffic rate, the Poisson distributed data traffic model had a mean arrival expectation of $1$ packet per second ($\lambda = 1.0$). As in the case with CBR, we computed the detection rate and the rate of false positives with the associated arithmetic averages and $95\%$ confidence  intervals. 

The results based on these two traffic models were similar, actually, we could not find the difference between them to be statistically significant. This points out that the detection process is robust against some variation in data traffic. This conclusion also reflects positively on the usefulness of the used genes. More importantly, it helped disperse our worries that the results presented in this experimental study could be unacceptably data traffic dependent.

\section{Related Work}
\label{sec:rel_work}

In~\cite{sarafijanovic2004ais,leboudec2004ais} the authors introduced an AIS based misbehavior detection system for ad hoc wireless networks. They used Glomosim for simulating data traffic, their setup was an area of 800$\times$600m with 40 mobile nodes (speed 1 m/s) of which 5-20 are misbehaving; the routing protocol was DSR. Four genes were used to capture local behavior at the network layer. The misbehavior implemented is a subset of misbehavior introduced in this paper; their observed detection rate is about 55\%. Additionally, a co-stimulation in the form of a danger signal was used in order to inform nodes on a forwarding path about misbehavior, thus propagating information about misbehaving nodes around the network.

In~\cite{hofmeyr1999ida} the authors describe an AIS able to detect anomalies at the transport layer of the OSI protocol
stack; only a wired TCP/IP network is considered. Self is defined as normal pairwise connections. Each detector is represented as a 49-bit string. The pattern matching is based on r-contiguous bits with a fixed $r=12$. 

Ref.~\cite{kim2001ens} discusses a network intrusion system that aims at detecting misbehavior by capturing TCP packet headers. 
They report that their AIS is unsuitable for detecting  anomalies in communication networks. This result is questioned in~\cite{balthrop2002rlp} where it is stated that this is due to the choice of problem representation and due to the choice of matching threshold $r$ for $r$-contiguous bits matching. 

To overcome the deficiencies of the generate-and-test approach a different approach is outlined in~\cite{kim2005dud}. Several signals each having a different function are employed in order to detect a specific misbehavior in sensor wireless networks. Unfortunately, no performance analysis was presented and the properties of these signals were not evaluated with respect to their misuse.

The main discerning factor between our work and works shortly discussed above is that 
we carefully considered hardware parameters of current sensor devices, the set of input parameters was designed in order to target specifically sensor networks  and our simulation setup reflects structural qualities of such networks with regards to existence of multiple independent routing paths. In comparison to~\cite{sarafijanovic2004ais,leboudec2004ais} we showed that in case of static sensor networks it is reasonable to expect the detection rate to be above 80\%.  

\section{Conclusions and Future Work}

Although we answered some basic question on the suitability and feasibility of AIS for detecting misbehavior in sensor networks a few questions remain open. 

The key question in the design of AIS is the quantity, quality and ordering of genes that are used for measuring behavior at nodes. To answer this question a detailed formal analysis of communications protocols will be needed. The set of genes should be as ``complete'' as possible with respect to any possible misbehavior. The choice of genes should impose a high degree of sensor network's survivability defined as �{\em the capability of a system to fulfill its mission in a timely manner, even in the presence of attacks, failures or accidents}~\cite{sterbenz2002smw}. 
It is therefore of paramount importance that the sensor network's mission is clearly defined and achievable under normal operating
conditions.

We showed the influence and usefulness of certain genes in order to detect misbehavior and the impact of the $r$ parameter on the detection process. In general, the results in Fig.~\ref{fig:det_used3} show that Gene \#1 and \#2 obtained of all genes the best results, with Gene \#2 showing always the best results. The contribution of Gene \#1 suggests that observing the MAC layer and the ratio of complete handshakes to the number of RTS packets sent is useful for the implemented misbehaviour. 

Gene \#2 fits perfectly for the implemented misbehavior. It therefore comes as no surprise that this gene showed the best results in the detection process.  The question which remains open is whether the two genes are still as useful when exposed to different attack patterns.  

It is currently unclear whether genes that performed well with negative selection, will also be appropriate for generating different flavors of signals as suggested within the {\em danger theory}~\cite{aickelin2003dtl,greensmith2005idc}. It is our opinion that any set of genes, whether used with negative selection or for generating any such a signal, should aim at capturing intrinsic properties of the interaction among different components of a given sensor network. This contradicts approaches applied in~\cite{sarafijanovic2004ais,kim2005dud} where the genes are closely coupled with a given protocol. The reason for this statement is the {\em combined performance} of Gene  \#1 and \#2. Their interaction can be understood as follows: data packet dropping implies less medium contention since there are less data packets to get forwarded. Less data packets to forward on the other hand implies easier access to the medium, i.e. the number of complete MAC handshakes should increase. This is an interesting {\em complementary} relationship since in order to deceive these two genes, a misbehaving node has to appear to be correctly forwarding data packets and, at the same time, he should not significantly modify the ``game'' of medium access. 

It is improbable that the misbehaving node {\em alone} would be able to estimate the impact of dropped packets on the contention level. Therefore, he lacks an important feedback mechanism that would allow him to keep the contention level unchanged. For that, he would need to act in collusion with other nodes. The property of complementarity moves the burden of excessive communication from normally behaving nodes to misbehaving nodes, thus, exploiting the ad hoc (local) nature of sensor networks. Our results thus imply, {\em a ``good'' mixture of genes should be able to capture interactions that a node is unable to influence when acting alone.} It is an open question whether there exist other useful properties of genes, other than complementarity.

We conclude that the random-generate-and-test process, with no knowledge of the used protocols and their behavior, creates many detectors which might show to be superfluous in detecting misbehavior. A process with some basic knowledge of protocol limitations might lead to improved quality of detectors.  

In~\cite{stibor2005nsa} the authors stated that the random-generate-and-test process {\em ``is innefficient, since a vast number of randomly generated detectors need to be discarded, before the required number of the suitable ones are obtained''.} Our results show that at $r = 10$, the rate of discarded detectors is less than $4\%$.
Hence, at least in our setting we could not confirm the above statement. A disturbing fact is, however, that the size of the self set in our setting was probably too small in order to justify the use of negative selection. A counter-balancing argument is here the realistic setup of our simulations and a decent detection rate.

We would like to point out that the Fisher iris and biomedical data sets, used in~\cite{stibor2005nsa} to argue about the apropriateness of negative selection for anomaly detection, could be very different from data sets generated by our simulations. Our experiments show that anomaly (misbehavior) data sets based on sensor networks could be in general very sparse. This effect can be due to the limiting nature of communications protocols. Since the Fisher iris and biomedical data sets were in~\cite{stibor2005nsa} not evaluated with respect to some basic properties e.g. degree of clustering, it is hard to compare our results with the results presented therein.

In order to understand the effects of misbehavior better (e.g. the propagation of certain adverse effects), we currently develop a general framework for AIS to be used within the JiST/SWANS network simulator~\cite{barr2005jea}. 

\section*{Acknowledgments}

This work was supported by the German Research Foundation (DFG) under the grant no. SZ 51/24-2 (Survivable Ad Hoc Networks -- SANE).

\balance
\bibliographystyle{plain}

\begin{thebibliography}{10}

\bibitem{aickelin2003dtl}
U.~Aickelin, P.~Bentley, S.~Cayzer, J.~Kim, and J.~McLeod.
\newblock {Danger Theory: The Link between AIS and IDS?}
\newblock {\em Proc. of International Conference on Artificial Immune Systems
  (ICARIS)}, pages 147--155, 2003.

\bibitem{aickelin4isa}
U.~Aickelin, J.~Greensmith, and J.~Twycross.
\newblock {Immune system approaches to intrusion detection - a review}.
\newblock {\em Proc. of International Conference on Artificial Immune Systems},
  pages 316--329, 2004.

\bibitem{bajaj1999gsn}
L.~Bajaj, M.~Takai, R.~Ahuja, K.~Tang, R.~Bagrodia, and M.~Gerla.
\newblock {GloMoSim: A Scalable Network Simulation Environment}.
\newblock {\em UCLA Computer Science Department Technical Report}, 990027,
  1999.

\bibitem{balthrop2002rlp}
J.~Balthrop, S.~Forrest, and M.~Glickman.
\newblock {Revisiting lisys: Parameters and normal behavior}.
\newblock {\em Proc. of Congress on Evolutionary Computation}, pages
  1045--1050, 2002.

\bibitem{banchereau2000idc}
J.~Banchereau, F.~Briere, C.~Caux, J.~Davoust, S.~Lebecque, Y.J. Liu,
  B.~Pulendran, and K.~Palucka.
\newblock {Immunobiology of dendritic cells}.
\newblock {\em Annual review of immunology}, 18(1):767--811, 2000.

\bibitem{barr2005jea}
R.~Barr, Z.J. Haas, and R.~van Renesse.
\newblock {JiST: an efficient approach to simulation using virtual machines}.
\newblock {\em Software Practice and Experience}, 35(6):539--576, 2005.

\bibitem{barrett2005upp}
C.~Barrett, M.~Drozda, DC~Engelhart, VSA Kumar, MV~Marathe, MM~Morin, SS~Ravi,
  and JP~Smith.
\newblock {Understanding protocol performance and robustness of ad hoc networks
  through structural analysis}.
\newblock {\em Proc. of the IEEE International Conference on Wireless And
  Mobile Computing, Networking And Communications (WiMob'2005)}, 3:65--72,
  2005.

\bibitem{cayzer2005hgl}
S.~Cayzer, J.~Smith, J.A.R. Marshall, and T.~Kovacs.
\newblock {What have Gene Libraries done for AIS?}
\newblock {\em Proc. of International Conference on Artificial Immune Systems
  (ICARIS)}, pages 86--99, 2005.

\bibitem{xbow}
{Crossbow Technologies Inc. {\tt www.xbow.com}}.

\bibitem{dasgupta2002ibt}
D.~Dasgupta and F.~Gonzalez.
\newblock {An immunity-based technique to characterize intrusions in computer
  networks}.
\newblock {\em IEEE Transactions on Evolutionary Computation}, 6(3):281--291,
  2002.

\bibitem{dhaeseleer1996iac}
P.~D'haeseleer, S.~Forrest, and P.~Helman.
\newblock {An Immunological Approach to Change Detection: Algorithms, Analysis
  and Implications}.
\newblock {\em IEEE Symposium on Security and Privacy}, pages 110--119, 1996.

\bibitem{drozda06smd}
M.~Drozda, S.~Schaust, and H.~Szczerbicka.
\newblock {Simulation of Misbehaviour Detection in Wireless Ad Hoc Networks}.
\newblock {\em Proc. of 19th Symposium on Simulation Technique (ASIM)}, pages
  235--240, 2006.

\bibitem{drozda2006ais}
M.~Drozda and H.~Szczerbicka.
\newblock Artificial immune systems: Survey and applications in ad hoc wireless
  networks.
\newblock {\em Proc. of International Symposium on Performance Evaluation of
  Computer and Telecommunication Systems (SPECTS'06)}, pages 485--492, 2006.

\bibitem{drozda2005aah}
M.~Drozda, H.~Szczerbicka, T.~Bessey, M.~Becker, and Barton R.
\newblock Approaching ad hoc wireless networks with autonomic computing: A
  misbehavior perspective.
\newblock {\em Proc. of International Symposium on Performance Evaluation of
  Computer and Telecommunication Systems (SPECTS'05)}, pages 723--733, 2005.

\bibitem{gonzalez2003ebm}
F.~Gonzalez, D.~Dasgupta, and J.~Gomez.
\newblock {The effect of binary matching rules in negative selection}.
\newblock {\em Proc. of Genetic and Evolutionary Computation Conference
  (GECCO)}, pages 196--206, 2003.

\bibitem{greensmith2005idc}
J.~Greensmith, U.~Aickelin, and S.~Cayzer.
\newblock {Introducing Dendritic Cells as a Novel Immune-Inspired Algorithm for
  Anomaly Detection}.
\newblock {\em Proc. of International Conference on Artificial Immune Systems
  (ICARIS)}, pages 153--167, 2005.

\bibitem{hofmeyr1999ida}
S.A. Hofmeyr and S.~Forrest.
\newblock {Immunity by design: An artificial immune system}.
\newblock {\em Proc. of Genetic and Evolutionary Computation Conference
  (GECCO)}, 2:1289--1296, 1999.

\bibitem{janewayjr2001isw}
C.A. Janeway~Jr.
\newblock {How the immune system works to protect the host from infection: A
  personal view}.
\newblock {\em Proc. of the National Academy of Sciences}, 98(13):7461--7468,
  2001.

\bibitem{ji2004rvn}
Z.~Ji and D.~Dasgupta.
\newblock {Real-valued negative selection algorithm with variable-sized
  detectors}.
\newblock {\em Proc. of Genetic and Evolutionary Computation Conference
  (GECCO)}, pages 287--298, 2004.

\bibitem{johnson1996dsr}
D.B. Johnson and D.A. Maltz.
\newblock {Dynamic source routing in ad hoc wireless networks}.
\newblock {\em Mobile Computing}, 353:153--181, 1996.

\bibitem{karl2005paa}
H.~Karl and A.~Willig.
\newblock {\em {Protocols and Architectures for Wireless Sensor Networks}}.
\newblock John Wiley and Sons, 2005.

\bibitem{kim2005dud}
J.~Kim, P.~Bentley, C.~Wallenta, M.~Ahmed, and S.~Hailes.
\newblock {Danger Is Ubiquitous: Detecting Malicious Activities in Sensor
  Networks Using the Dendritic Cell Algorithm}.
\newblock {\em Proc. of International Conference on Artificial Immune Systems
  (ICARIS)}, pages 390--403, 2006.

\bibitem{kim2001ens}
J.~Kim and P.J. Bentley.
\newblock {An evaluation of negative selection in an artificial immune system
  for network intrusion detection}.
\newblock {\em Proc. of Genetic and Evolutionary Computation Conference
  (GECCO)}, pages 1330--1337, 2001.

\bibitem{leboudec2004ais}
J.Y. Le~Boudec and S.~Sarafijanovic.
\newblock {An Artificial Immune System Approach to Misbehavior Detection in
  Mobile Ad-Hoc Networks}.
\newblock {\em Proc. of Bio-ADIT}, pages 96--111, 2004.

\bibitem{marti2000mrm}
S.~Marti, TJ~Giuli, K.~Lai, and M.~Baker.
\newblock {Mitigating routing misbehavior in mobile ad hoc networks}.
\newblock {\em Proc. of International Conference on Mobile Computing and
  Networking}, pages 255--265, 2000.

\bibitem{sarafijanovic2004ais}
S.~Sarafijanovic and J.Y. Le~Boudec.
\newblock {An artificial immune system for misbehavior detection in mobile
  ad-hoc networks with virtual thymus, clustering, danger signal and memory
  detectors}.
\newblock {\em Proc. of International Conference on Artificial Immune Systems
  (ICARIS)}, pages 342--356, 2004.

\bibitem{sterbenz2002smw}
J.P.G. Sterbenz, R.~Krishnan, R.R. Hain, A.W. Jackson, D.~Levin, R.~Ramanathan,
  and J.~Zao.
\newblock {Survivable mobile wireless networks: issues, challenges, and
  research directions}.
\newblock {\em Proc. of ACM workshop on Wireless security}, pages 31--40, 2002.

\bibitem{stibor2005nsa}
T.~Stibor, P.~Mohr, J.~Timmis, and C.~Eckert.
\newblock {Is negative selection appropriate for anomaly detection?}
\newblock {\em Proc. of Conference on Genetic and evolutionary computation},
  pages 321--328, 2005.

\bibitem{ye2004mac}
W.~Ye and J.~Heidemann.
\newblock {Medium Access Control in Wireless Sensor Networks}.
\newblock {\em Wireless Sensor Networks}, pages 73--91, 2004.

\bibitem{zhang2003idt}
Y.~Zhang, W.~Lee, and Y.A. Huang.
\newblock {Intrusion Detection Techniques for Mobile Wireless Networks}.
\newblock {\em Wireless Networks}, 9(5):545--556, 2003.

\end{thebibliography}

\end{document}